\documentclass{article}

\usepackage{arxiv}

\usepackage[utf8]{inputenc} 
\usepackage[T1]{fontenc}    
\usepackage{hyperref}       
\usepackage{url}            
\usepackage{booktabs}       
\usepackage{amsfonts}       
\usepackage{nicefrac}       
\usepackage{microtype}      
\usepackage{lipsum}		
\usepackage{doi}
\usepackage{parskip}
\usepackage{color}
\usepackage{caption}
\usepackage{subcaption}
\usepackage{float}
\usepackage{mathematics}
\usepackage{graphics,graphicx}
\usepackage{epstopdf}
\usepackage{wrapfig}
\usepackage{amsmath}
\usepackage{cancel}
\usepackage{mathabx}
\usepackage{enumitem}
\usepackage{todonotes}
\usepackage{multirow}
\usepackage{xcolor}

\graphicspath{{Figures/}}

\usepackage{soul}

\newcommand{\Added}[1]{#1}

\title{\Added{Identifiability and Characterization of Transmon Qutrits Through Bayesian Experimental Design}}

\author{ 
\href{https://orcid.org/0000-0001-6882-9737}{\includegraphics[scale=0.06]{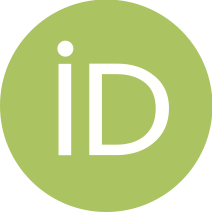}\hspace{1mm}
    Sohail Reddy}\thanks{Corresponding author.} \\
	Center for Applied Scientific Computing \\
	Lawrence Livermore National Laboratory \\
	Liveremore, CA 94550 \\
	\texttt{reddy6@llnl.gov} \\
}

\renewcommand{\shorttitle}{Identifiability and Bayesian Quantum Characterization}
\newcommand{\etal}{\textit{et al.}}

\begin{document}
\maketitle

\begin{abstract}

	Robust control of a quantum system is essential to utilize the current noisy quantum hardware to their full potential, such as quantum algorithms. To achieve such a goal, systematic search for an optimal control for any given experiment is essential. Design of optimal control pulses require accurate numerical models, and therefore, accurate characterization of the system parameters. We present an online, Bayesian approach for quantum characterization of qutrit systems which automatically and systematically identifies the optimal experiments that provide maximum information on the system parameters, thereby greatly reducing the number of experiments that need to be performed on the quantum testbed. Unlike most characterization protocols that provide point-estimates of the parameters, the proposed approach is able to estimate their probability distribution. The applicability of the Bayesian experimental design technique was demonstrated on test problems where each experiment was defined by a parameterized control pulse. In addition to this, we also presented an approach for \Added{iterative pulse extension} which is robust under uncertainties in transition frequencies and coherence times, and shot noise, despite being initialized with wide uninformative priors. Furthermore, we provide a mathematical proof of the theoretical identifiability of the model parameters and present conditions on the quantum state under which the parameters are identifiable. The proof and conditions for identifiability are presented for both closed and open quantum systems using the Schroedinger equation and the Lindblad master equation respectively. 
	
\end{abstract}


\section{Introduction} \label{sec:Intro}

The Noisy Intermediate Scale Quantum (NISQ) devices are poised to demonstrate quantum advantage \cite{Akhalwaya:2022}. They are, however, very sensitive to a myriad of noise sources such as from interaction with the environment, characterized by coherence times \cite{Gao:2007,Klimov:2018}, and fluctuating quantum systems and interaction with defects \cite{Wang:2015,Cho:2023}, causing uncertainties in qubit transition frequencies. This instability requires quantum error correction and dynamical decoupling techniques which rely on a larger number of qubits and lengthy circuit depth to encode quantum information and cancel out the noises \cite{Roffe:2019}, thereby limiting the quantum algorithms that can be implemented within the device's coherence times. Hence, variational quantum algorithms (VQA) \cite{Cerezo:2021,Ravi:2022}, which couple computations on the stable classical hardware with the noisy quantum processing units (QPUs), have emerged as a promising strategy to address these drawbacks. Such classical-quantum algorithms typically require accurate modeling of the quantum systems. These numerical models can be parameterized by a few quantities of the QPUs, thus, accurate determination of these parameters is paramount for high-fidelity simulation of the QPU. However, the noisy behavior of current devices poses a great challenge to identifying the testbed characteristics, particularly, providing point-estimates of these parameters. Under such circumstance, posing the characterization problem within a probabilistic Bayesian framework is ideal. Regardless of the approach used for characterization, the identifiability of the model parameters must first be established. Although crucial, such \emph{priori} analysis is often neglected. Such identifiability analysis not only determine the parameters that can reliably and uniquely estimated, but can also identify the correlated parameters. Furthermore, they establish a set of conditions on the experiments/measurements that make parameters identifiable, thereby informing the design of experiments. One such approach involves characterization (i.e., parameter estimation) through online Bayesian experimental design (BExD). The BExD technique uses classical optimization techniques to identify experimental parameters (or experiments) that provide maximum \emph{information} about the parameters being identified and iteratively updates the prior distribution of the parameters. Estimating complete probability distributions provide insight on the correlation between parameters and allow for design of control pulses that are robust to slight variances in parameters. Furthermore, such distributions can better inform development of noise models and has been an active area of research in the NISQ era \cite{Magesan:2013,Georgopoulos:2021,Onorati:2023}.

The BExD approach has been used to identify parameters in electro-chemical models of Lithium-Ion cells \cite{Streb:2022}, pharmacokinetic model \cite{Ryan:2015}, Thomson scattering model for nuclear fusion \cite{Fischer:2004} and electrical impedance tomography \cite{Karimi:2021}. The probabilistic nature of quantum mechanics naturally lends itself to Bayesian characterization. The BExD technique has been used to estimate the resonant frequency of a single two-level closed system under time-independent Hamiltonian by minimizing the Bayes' risk \cite{Ferrie:2013}. Granade \etal \cite{Granade:2012} demonstrated the BExD approach to infer the transition frequency and $T_2$ dephasing term on a canonical, two-level system, and built the open-source framework QInfer \cite{Granade:2017}. McMichael \etal \cite{McMichael:2021} utilized BExD to determine optimal controls in Ramsey interferometry to estimate precession frequency and $T_2^\star$ decoherence time, and demonstrated a five-factor speed up over random sampling methods. Wang \etal \cite{Wang:2022} demonstrated the BExD framework for quantum sensing of the environment around a Nitrogen-Vacancy (NV) center in diamond and reported a 90\% speed up over their heuristic method in estimating the nearby nuclear spins and oscillating magnetic field. Hincks \etal \cite{Hincks:2018} estimated five-parameters in their Hamiltonian and Lindblad model for NV in diamonds with Ramsey and Rabi style experiments where they showed the BExD approach results in a posterior variance that is two orders of magnitude smaller than those obtained using the heuristic methods. Wang \etal \cite{Wang:2017} performed characterization of an NV center in diamonds by estimating the Rabi frequency and updating their Hamiltonian model to include missing interactions (chirping). Both Wang \etal \cite{Wang:2017} and Ferrie \etal \cite{Ferrie:2013} demonstrated exponential convergence of the BExD (when minimizing posterior variance) up to a saturation point, beyond which the ansatz is no longer able to model the missing interactions. \Added{Gester \etal \cite{Gerster2022} recently employed the BExD approach to calibrate two-qubit entangling operators for a closed trapped-ion systems using $1200 \pm 500$ experiments in less than one minute. Although the BExD approach significantly reduced the number of experiments required for accurate estimation of the model parameters, the significant speed up is mainly due to the use of analytical solution to their Hamiltonian model. Stace \etal  \cite{Stace2024} demonstrated the BExD technique on single qubit and synthetic two qubit trapped-ion systems, where they estimated the detuning frequency, Rabi frequency and the coupling strength using optimized bang-bang pulses. They reported difficulty in estimating model parameters where the algorithm fails to converge which is attributed to the identifiability of the parameter due to the states prepared under the pulse parameterization.} Although paramount, most studies for characterization of the quantum system often do not consider the identifiability of the parameters under a given experimental setup. Wang \etal \cite{Wang2020} demonstrated that the identifiability of closed quantum systems with a single qubit in density matrix formalism (Liouville-von Neumann equation). Zhang \etal \cite{Zhang2009} investigated the identifiability of both closed and open, dephasing qubit systems and demonstrated the identifiability can be greatly increased by performing appropriate experiments that can be parameterized by control pulses driving the quantum system.

However, the identifiability of multi-level closed and open quantum systems has yet to be explored. Hence, this work demonstrates the identifiability of parameters in the Schroedinger and Lindblad models of multi-level closed and open (with energy decay and dephasing) systems, respectively, and the sufficient conditions on quantum state for the parameters to be identifiable under a set of measurements. Furthermore, it demonstrates the invariance of the conditions to unitary transformations, hence their applicability to a different set of measurements. The BExD approach is then used to identify optimal experiments. Identifying these optimal experimental parameters amounts to maximizing functions of the experiment controls, parameters, and prior that quantifies how informative the experiment would be if performed. These utility functions will be discussed in the following sections. Although demonstrated on a single, qubit problems, its performance for characterizing quantum systems with higher energy levels has yet to be thoroughly investigated. This work aims to address this scarcity and apply the BExD technique for characterizing multi-level quantum systems. We present an approach for \Added{iterative pulse extension} to accurately estimate parameters that govern processes operating on different time scales.

\section{Quantum Mechanical Model} \label{sec:Quantum:Model}

Consider an open quantum system whose dynamics are modeled by the Lindblad master equation \cite{Breuer2007}
\eq{Quantum:MasterEq}{
    \ddt{\rho} = -i \sbrac{H, \rho} + L(\rho)
}
where $\rho \in \C^{3 \times 3}$ is the density matrix, $[\cdot,\cdot]$ is the commutator, $H(t)\in \C^{3 \times 3}$ is the Hamiltonian governing the unitary dynamics, $L(\rho)$ is the Lindbladian governing the dissipative dynamics. The Hamiltonian in the lab frame consists of the time-independent system Hamiltonian $H_s$ and a time-dependent control Hamiltonian $H_c(t)$
\aligneq{Quantum:Hamiltonian}{
    H(t) := \underbrace{\omega  a^\dagger a - \dfrac{\chi}{2}a^\dagger a^\dagger a a}_{H_s} + \underbrace{ f(t) (a + a^\dagger) }_{H_c(t)} \\
}
where $a$ and $a^\dagger$ are the lowering and raising operators, $\omega$ is the 0-1 transition frequency, $\chi$ is the anharmonicity, and $f(t) = 2 \Re(\Omega(t)e^{iw_d t})$ is the control function with drive frequency $\omega_d$ and amplitude $\Omega(t):= p(t) + i q(t)$ where $p(t), q(t) \in \R$. Under the rotating wave approximation (RWA), we obtain the Hamiltonian in the rotating frame
\aligneq{Quantum:RWAHamiltonian}{
    H(t) &= \underbrace{ \rbrac{\omega - \omega_d} a^\dagger a - \dfrac{\chi}{2}a^\dagger a^\dagger a a}_{H_s} + \underbrace{ \Omega(t) a + \overline{\Omega}(t)a^\dagger }_{H_c(t)} \\
    &= \Delta a^\dagger a - \dfrac{\chi}{2}a^\dagger a^\dagger a a + p(t) \rbrac{ a + a^\dagger} + i q(t) \rbrac{ a - a^\dagger}
}
where the rotation frequency in the RWA is taken to be the drive frequency $\omega_d$, and $\Delta := \omega - \omega_d$ is the detuning between the qubit 0-1 transition and the drive frequencies. \Added{This model was experimentally validated for a transmon qutrit at Quantum Device and Integration Testbed (QuDIT) at LLNL \cite{Cho2023:1}.} The decoherence is introduced through the Lindblad jump operators for decay ($L_{1} := \frac{1}{\sqrt{T_1}}a$) and dephasing ($L_{2} := \frac{1}{\sqrt{T_2}} a^\dagger a$) where $T_1$ and $T_2$ are the decay and dephasing times, respectively. The Lindbladian is then written as
\eq{Quantum:Lindbladian}{
    L(\rho) := \sum_{i=1}^2 L_{i} \rho L^\dagger_{i} - \dfrac{1}{2} \cbrac{ L^\dagger_{i}  L_{i},~\rho}
}
where $\cbrac{\cdot,\cdot}$ is the anti-commutator. The control pulse is modeled as a set of piece-wise constant pulses, each with $p(t)$, $q(t)$, and duration $\Delta t$. The pulse resolution can be increased by increasing the number of segments. In general, real experiments require 1 ns resolution to play pulses on hardware. This generalization can be seen as a bang-bang control where each application uses a different amplitude and duration for the control. Equation \ref{Eq:Quantum:MasterEq} is vectorized 
\expression{\ddt{\tilde{\rho}} = \mathcal{L}(t)\tilde{\rho}}
with $\tilde{\rho} = \mathrm{vec}(\rho)$ and evolved until final time $T$ using matrix exponentiation by means of the Padé approximant
\eq{Quantum:Solve}{
    \tilde\rho(T) = \prod_{i=1}^{N_T} \exp( \mathcal{L}(t_i) \Delta t_i) \tilde\rho(0)
}
where $N_T$ is the number of piecewise-constant segments and the Liouvillian superoperator $\mathcal{L}(t)$ is defined as
\eq{Quantum:Superoperator}{
    \mathcal{L}(t) := -i\rbrac{I \otimes H(t) - H^\dagger(t) \otimes I} + \sum_{i=1}^2 L_{i} \otimes L^\dagger_{i} - \dfrac{1}{2} \rbrac{ I \otimes L^\dagger_{i}  L_{i} + L^\dagger_{i}  L_{i} \otimes I}
}
and $I$ is rank-3 identity matrix. Given a density matrix, the probability of measuring the system in a particular state $y$ is given as
\eq{Quantum:Measurement}{
    \Pr{y|\rho} = \Tr{M_y \rho}
}
where $M_y$ is the measurement operator corresponding to the observable $y$. In this work, we assume projective measurements in the $z$-basis and take $M_y = \ketbra{e_y}{e_y},~y=0,1,2$ where $\ket{e_i}$ is the $i^{th}$ canonical unit vector; then each element of $\Pr{\Vec{y}|\rho} = \mathrm{diag}(\rho)$ denotes the probability of measuring that state and will be used to compute the likelihood in the subsequent sections.

\section{Identifiability of Quantum Mechanical Models} \label{sec:Quantum:Identifiability}

Before characterizing a system described by a mathematical model, it is imperative to determine the global and unique (termed \emph{global structural}) identifiability of the parameters in such a model. Some techniques for determining the identifiability of a model include those based on Laplace transforms \cite{Bellman1970}, similarity transformation \cite{Vajda1989}, generating series \cite{Walter1982}, differential algebra \cite{Ljung1994,Bellu2007} and the Taylor series approach \cite{Pohjanpalo1978}. To determine the identifiability of the models of open and closed quantum systems, we will employ the Taylor series approach \cite{Pohjanpalo1978,Mouroutsos1985,Chis2011}. The approach relies on the observations being a unique analytic function of time, hence their derivatives must also be unique. The uniqueness of this representation will guarantee the structural identifiability of the model. Consider an observable/measurable quantity $y(t;\Vec{\theta})$ that is dependent on model parameters $\Vec{\theta}$. The Taylor series approach expands this quantity about time $t_0$
\eq{Identifiability:TSE}{
    y_i\rbrac{t;\Vec{\theta}} \approx \sum_{k=0}^{m} \dfrac{t^k}{k!} a_{i,k}(\Vec{\theta}), \quad \mathrm{with}~a_{i,k} = \Tddt[^k]{}y_i\rbrac{t_0;\Vec{\theta}}
}
A sufficient condition for global structural identifiability is given by the uniqueness of the expansion coefficients
\eq{Identifiability:GSI}{
    a_{i,k}(\Vec{\theta}) = a_{i,k}(\Vec{\theta}^\star),~~ k=1,...,m; i=0,1,2 \Rightarrow \Vec{\theta}=\Vec{\theta}^\star
}
where $m$ is the smallest positive integer such that the resulting system is solvable for the parameters. \Added{We emphasize that the order of expansion here is used only to establish identifiability conditions.} This minimum order of expansion is not known a \textit{priori} but as will be shown, first order and second order expansions are sufficient for closed and open quantum systems, respectively. \Added{We should mention that the identifiability conditions are derived for models in the lab-frame and, therefore, do not make any approximations such as neglecting the rapidly oscillating terms in the RWA.}

\subsection{Model Identifiability: Schroedinger Equation} \label{subsec:Quantum:Identifiability:Schroedinger}

Consider a closed quantum system modeled by the Schroedinger equation $\dot{\ket{\psi}} = -i H \ket{\psi}$, driven by a \emph{known} pulse $f(t)$ with a pulse amplitude $\Omega(t)$ and the probability of measuring state $\ket{i}$ by $y_i(t;\Vec{\theta}):=\Pr{\ket{i},t;\Vec{\theta}}=\bra{\Vec{\psi}(t;\Vec{\theta})}M_i\ket{\Vec{\psi}(t;\Vec{\theta})}$. \Added{The pulse amplitude $\Omega(t)$ is taken to be constant for all times. An alternative parameterization for $\Omega(t)$ can also be used, in which case, the resulting Taylor series expansion will yield additional pulse dependent terms that simply scale the coefficients $a_{i,k}$ in \eqref{Identifiability:TSE} and, therefore, do not alter the identifiability conditions.} We expand the measurement about $t_0=0$ with $\mathcal{F} := f(t_0)$, which yields the first three expansion coefficients
\aligneq{Quantum:Identifiability:Coeff}{
    a_{i,0}(\Vec{\theta}) &= \bra{\Vec{\psi}(t_0)} M_i \ket{\Vec{\psi}(t_0)}\\
    a_{i,1}(\Vec{\theta}) &= i\bra{\Vec{\psi}(t_0)} \sbrac{H,M_i}\ket{\Vec{\psi}(t_0)} \\
    a_{i,2}(\Vec{\theta}) &= \bra{\Vec{\psi}(t_0)} \sbrac{\sbrac{H,M_i},H}\ket{\Vec{\psi}(t_0)} \\
}
We note the zeroth order terms ($a_{i,0}(\Vec{\theta}),~i=0,1,2$) do not yield any information on the parameters of the model. Defining $\Vec{\psi}(t_0) = \cbrac{\psi_0, \psi_1, \psi_2} \in \mathbb{C}^3$, the first order terms can be written as 
\aligneq{Quantum:Identifiability:LinearCoeff}{
    a_{0,1}(\Vec{\theta}) &= 2\mathcal{F}~\Im(\overline{\psi_0}\psi_1) \\
    a_{1,1}(\Vec{\theta}) &= 2\mathcal{F} \rbrac{\Im(\psi_0 \overline{\psi_1}) + \sqrt{2}\Im(\overline{\psi_1}\psi_2)} \\
    a_{2,1}(\Vec{\theta}) &= 2\sqrt{2}\mathcal{F}~\Im(\psi_1 \overline{\psi_2})
}
which uniquely define $\mathcal{F}$ when $\Im(\psi_i \overline{\psi_j}) \neq 0$ for $i\neq j$. The second order terms are given by
\aligneq{Quantum:Identifiability:QuadraticCoeff}{
    a_{0,2}(\Vec{\theta}) &= 2\mathcal{F}^2 \rbrac{\abs{\psi_1}^2 - \sqrt{2}\Re(\psi_0 \overline{\psi_2}) - \abs{\psi_0}^2} - 2 \omega \mathcal{F} \rbrac{ \Re(\psi_0 \overline{\psi_1}) } \\
    a_{1,2}(\Vec{\theta}) &= 2\mathcal{F}^2 \rbrac{\abs{\psi_0}^2 - 3\abs{\psi_1}^2 + 2\abs{\psi_2}^2 + 2\sqrt{2}\Re(\psi_0 \overline{\psi_2})} + 2 \omega \mathcal{F} \rbrac{ \Re(\psi_0 \overline{\psi_1}) - \sqrt{2} \Re(\psi_1 \overline{\psi_2}) } \\
    & + 2 \sqrt{2} \chi \mathcal{F} \rbrac{ \Re(\psi_1 \overline{\psi_2}) } \\
    a_{2,2}(\Vec{\theta}) &= 2\mathcal{F}^2 \rbrac{ 2\abs{\psi_1}^2 - 2\abs{\psi_2}^2 - \sqrt{2}\Re(\psi_0 \overline{\psi_2})} + 2 \sqrt{2} \omega \mathcal{F} \rbrac{  \Re(\psi_1 \overline{\psi_2}) } - 2 \sqrt{2} \chi \mathcal{F} \rbrac{ \Re(\psi_1 \overline{\psi_2}) }
}
Collecting all the known and identified terms in a right-hand-side $\mathcal{S} \in \R^2$, we can write the identifiability conditions in matrix form
\aligneq{Quantum:Identifiability:LinearCoeffMatrix}{
\underbrace{\rmatrix{ \Re(\psi_0 \overline{\psi_1}) & 0  \\
\Re(\psi_0 \overline{\psi_1}) - \sqrt{2} \Re(\psi_1 \overline{\psi_2}) & \sqrt{2} \Re(\psi_1 \overline{\psi_2})  \\
                      \Re(\psi_1 \overline{\psi_2}) & -\Re(\psi_1 \overline{\psi_2})
            }}_{\Added{A_c}} \cmatrix{\omega \\ \chi } = \mathcal{S} 
}
Notice that the rows \Added{of $A_c$} are linearly dependent. Since the drive $\mathcal{F}$ is known (or can be identified as indicated by the first order terms), the sufficient conditions to identify $\omega$ and $\chi$ are $\Re(\psi_0 \overline{\psi_1}) \neq 0$ and $\Re(\psi_1 \overline{\psi_2}) \neq 0$, respectively. Furthermore, the sufficient conditions for identifiability of $\chi$ requires that the conditions for $\omega$ also be satisfied. Hence, the parameters are globally and uniquely identifiable and the model is \emph{globally structrually identifiable}, when the control pulse is able to prepare a state that satisfies these conditions.

\subsection{Model Identifiability: Lindblad Master Equation} \label{subsec:Quantum:Identifiability:Lindblad}

We now determine the identifiability of the Lindblad master equation. Consider an open quantum system defined by \eqref{Quantum:MasterEq} with observables defined as $y_i(t;\Vec{\theta}) = \Tr{M_i \rho(t;\Vec{\theta})}$ where $M_i$ is a measurement operator defined in Sec. \ref{sec:Quantum:Model}.  As before, we let $\mathcal{F} := f(t_0)$ and expand the observable about $t_0=0$ (\eqref{Identifiability:TSE}) with coefficients

\aligneq{Append:Quantum:Identifiability:Coeff}{
    a_{i,0}(\Vec{\theta}) &= \Tr{M_i \rho(t_0)}\\
    a_{i,1}(\Vec{\theta}) &= -i\Tr{M_i \sbrac{H(t), \rho(t_0)}} + \sum_{k=1}^2 \Tr{M_i \rbrac{L_{k} \rho(t_0) L^\dagger_{k} - \dfrac{1}{2} \cbrac{ L^\dagger_{k}  L_{k},~\rho(t_0)}}} \\
    a_{i,2}(\Vec{\theta}) &= -i\Tr{M_i \sbrac{H(t), \dot{\rho}(t_0)}} + \sum_{k=1}^2 \Tr{M_i \rbrac{L_{k} \dot{\rho}(t_0) L^\dagger_{k} - \dfrac{1}{2} \cbrac{ L^\dagger_{k}  L_{k},~\dot{\rho}(t_0)}}} 
}

The second order terms $a_{i,2}(\Vec{\theta})$ are obtained by substituting the right-hand side of \eqref{Quantum:MasterEq} into $a_{i,2}(\Vec{\theta})$. Letting 
\eq{Appen:Quantum:Identifiability:DensityMatrix}{
    \rho_0 := \rho(t_0) = \rmatrix{ \rho_{00} & \overline{\rho_{10}} & \overline{\rho_{20}} \\
                     \rho_{10} & \rho_{11} & \overline{\rho_{21}} \\
                     \rho_{20} & \rho_{21} & \rho_{22} }
}
Letting $\tau_1 = \frac{1}{T_1}$ and $\tau_2 = \frac{1}{T_2}$, the first order terms can be expanded as
\aligneq{Quantum:Identifiability:LinearCoeff}{
    a_{0,1}(\Vec{\theta}) &= \tau_1 \rho_{11} - i\mathcal{F}\rbrac{\rho_{10} - \overline{\rho_{10}}} = \tau_1 \rho_{11} + 2 \mathcal{F}~\Im(\rho_{10})\\
    a_{1,1}(\Vec{\theta}) &=  \tau_1 \rbrac{2\rho_{22} - \rho_{11}} + i\mathcal{F}\rbrac{\rho_{10} - \overline{\rho_{10}} + \sqrt{2}\rbrac{\overline{\rho_{21}} - \rho_{21}} }\\
    &= \tau_1 \rbrac{2\rho_{22} - \rho_{11}} + 2\mathcal{F}\rbrac{\sqrt{2}\Im(\rho_{21}) - \Im(\rho_{10})}\\
    a_{2,1}(\Vec{\theta}) &= -2 \tau_1 \rho_{22} + i \sqrt{2} \mathcal{F} \rbrac{ \rho_{21} - \overline{\rho_{21}}} =  -2 \tau_1 \rho_{22} - 2\sqrt{2} \mathcal{F}~\Im(\rho_{21})
}
Note that $a_{2,1}(\Vec{\theta}) = -(a_{0,1}(\Vec{\theta}) + a_{1,1}(\Vec{\theta}))$. We can construct the linear system using $a_{0,1}(\Vec{\theta})$ and $a_{2,1}(\Vec{\theta})$
\aligneq{Append:Quantum:Identifiability:FirstOrderLinearCoeffMatrix}{
\underbrace{\rmatrix{ 2\Im(\rho_{10}) & \rho_{11} \\
                        \sqrt{2}\Im(\rho_{21}) & \rho_{22}
            }}_{\Added{A_1}} \cmatrix{\mathcal{F} \\ \tau_1 } = \mathcal{S}_1
}
Hence, if \Added{$A_1$} is non-singular, the first order terms prove global and unique identifiability of $\mathcal{F}$ and $\tau_1$. The second order terms are given as 
\aligneq{Quantum:Identifiability:QuadraticCoeff}{
    a_{0,2}(\Vec{\theta}) =& -2\omega\mathcal{F} ~\Re(\rho_{10}) + \tau_1\rbrac{\rho_{11} + \mathcal{F}\rbrac{2\sqrt{2}\Im(\rho_{21}) - \Im(\rho_{10}) }} - \tau_2\mathcal{F}~\Im(\rho_{10}) \\
     & + 2\mathcal{F}^2\rbrac{\rho_{11} - \rho_{00} - \sqrt{2}\Re(\rho_3)} \\
    a_{1,2}(\Vec{\theta}) =& 2\omega\mathcal{F}~\rbrac{\Re(\rho_{10})-\sqrt{2}\Re(\rho_{21})} + 2\sqrt{2}\mathcal{F}~\chi \Re(\rho_{21}) \\
    & + \tau_1\rbrac{2\rho_{22} +\tau_1\rbrac{\rho_{11} -2 \rho_{22}} +\mathcal{F}\rbrac{3\Im(\rho_{10}) - 7\sqrt{2}\Im(\rho_{21})}}\\
    & + \tau_2\underbrace{\rbrac{\rho_{11} - \tau_1\rbrac{2\rho_{22} - \rho_{11}} + \mathcal{F} \rbrac{3\Im(\rho_{10}) - 3\sqrt{2}\Im(\rho_{21})}}}_{A_{12}}\\
    & + 2\mathcal{F}^2\rbrac{\rho_1 - 3\rho_{11} + 2\rho_{22} + 2\sqrt{2}\Re(\rho_{20})}\\
    a_{2,2}(\Vec{\theta}) =& 2\sqrt{2}\omega \mathcal{F}~\Re(\rho_{21}) - 2\sqrt{2}\mathcal{F}~\chi\Re(\rho_{21}) + \tau_1 \rbrac{4\tau_1\rho_{22} + 7\sqrt{2} \mathcal{F}~\Im(\rho_{21})} \\
    & + \tau_2 \underbrace{\rbrac{4 \rho_{22} \rbrac{1+2\tau_1} +  9\sqrt{2}\mathcal{F}~\Im(\rho_{21})}}_{A_{22}} + 2\mathcal{F}^2\rbrac{2\rho_{11} - 2\rho_{22} - \sqrt{2}\Re(\rho_{20})}
}
Since $\mathcal{F}$ and $\tau_1$ are identifiable, we can construct the linear system
\aligneq{Append:Quantum:Identifiability:LinearCoeffMatrix}{
\underbrace{\rmatrix{ 2\mathcal{F}\Re(\rho_{10}) & 0 & -\mathcal{F}\Im(\rho_{10}) \\
                        2\mathcal{F} \rbrac{\Re(\rho_{10})-\sqrt{2}\Re(\rho_{21})} & 2\sqrt{2}\mathcal{F} \Re(\rho_{21})& A_{12} \\
                        2\sqrt{2}\mathcal{F} \Re(\rho_{21}) & -2\sqrt{2}\mathcal{F} \Re(\rho_{21}) & A_{22}
            }}_{\Added{A_2}} \cmatrix{\omega \\ \chi \\ \tau_2 } = \mathcal{S}_2 
}
for the remaining unknown parameters where the remaining terms involving the known or identifiable parameters ($\mathcal{F}$ and $\tau_1$) are grouped in $\mathcal{S}_2 \in \R^3$. These second order terms show that $\omega,~\chi,~\tau_2$ are globally identifiable if \Added{$A_2$} is non-singular. Hence, the control function should be able to drive the system to such a state. Similarly, the identifiability of the density matrix formalism for closed system (i.e., Liouville-von Neumann equation) can be determined by fixing $\tau_1 = \tau_2 = 0$. In this case, the first order conditions only show identifiability of $\mathcal{F}$, like in the case of state vector formalism (i.e., Schroedinger equation \S\ref{subsec:Quantum:Identifiability:Schroedinger}), and a second order expansion is needed to determine identifiability of the remaining parameters $\omega$, $\chi$. Note that these identifiability conditions in both density matrix and state vector formalism are equivalent. Letting $\rho(t_0) = \ketbra{\Vec{\psi}(t_0)}{\Vec{\psi}(t_0)}$, we obtain the condition for $\Im(\rho_{ij}) = \Im(\psi_i \overline{\psi_j}) \neq 0$ for $i \neq j$ for parameter $\mathcal{F}$ , $\Re(\rho_{01}) = \Re(\psi_0 \overline{\psi_1}) \neq 0$ for parameter $\omega$ and $\Re(\rho_{21}) = \Re(\psi_2 \overline{\psi_1}) \neq 0$ for parameter $\chi$. We should mention that these conditions also hold for other measurement basis $\widetilde{M}$ by applying an appropriate unitary transformation $U$ such that $\widetilde{M}_i = U M_i U^\dagger$ since $y_i(t) = \Tr{U M_i U^\dagger \rho(t)} = \Tr{M_i U^\dagger \rho(t) U}= \Tr{M_i  \widetilde{\rho}(t)}$ where the conditions on the quantum state now apply in the transformed basis. These conditions show that the model parameters are identifiable by measuring the population in a single basis without needing to estimate the full density matrix via state or process tomography, which can be expensive.

\section{Bayesian Experimental Design} \label{sec:BExD}

Having demonstrated the identifiability of the model parameters is dependent on the quantum state, and therefore, the control applied (or experiment performed), we present an approach for automatic, online design of experiments that allow the parameters to be identified. Consider a set of model parameters $\Vec{\theta}$ to be identified, set of measurements $\Vec{y}$ from an experiment parameterized by a set of controls (i.e. controllable parameters) $\Vec{\xi}$. We begin with the Bayes' theorem
\eq{Bayes}{
    \Pr{\Vec{\theta} | \Vec{y}; \Vec{\xi}} = \dfrac{\Pr{\Vec{y} | \Vec{\theta}; \Vec{\xi}} \Pr{\Vec{\theta} }}{\Pr{ \Vec{y} | \Vec{\xi}}}
}
where $\Pr{\Vec{\theta}}$, $\Pr{\Vec{\theta} | \Vec{y}; \Vec{\xi}}$, and $\Pr{\Vec{y} | \Vec{\theta}; \Vec{\xi}}$ are the prior, posterior distribution and likelihood, respectively. The normalizing constant is the marginalized probability distribution over parameter space (also referred to as the evidence) and is defined as 
\eq{Bayes:Marginalized}{
    \Pr{ \Vec{y} | \Vec{\xi}} = \integral{\Vec{\theta}}{}{\Pr{\Vec{y} | \Vec{\theta}; \Vec{\xi}} \Pr{\Vec{\theta} }~d\Vec{\theta}}
}
The Bayesian experimental design (BExD) framework aims to estimate the probability distributions of the model parameters (i.e. the posterior distribution) by identifying the optimal controls of an experiment that provides maximum information on the parameters. The measure of effectiveness of the experiment (parameterized by $\Vec{\xi}$) if the outcome $y$ is observed is called the utility function $U(y | \Vec{\xi})$. Hence, an optimal experiment is one that maximizes the expectation of this utility function over all possible observations
\aligneq{BExD:MarUtil}{
    U(\Vec{\xi}) = \E[{\Pr{ \Vec{y} | \Vec{\xi}}}]{U(\Vec{y} |\Vec{\xi})} & = \integral{\Vec{y}}{}{ U(\Vec{y} | \Vec{\xi}) \Pr{ \Vec{y} | \Vec{\xi}} ~d\Vec{y}}\\
    & = \integral{\Vec{\theta}}{}{\integral{\Vec{y}}{}{ U(\Vec{y} | \Vec{\xi}) \Pr{\Vec{y} | \Vec{\theta}; \Vec{\xi}} \Pr{\Vec{\theta} } ~d\Vec{y}} ~d\Vec{\theta}}
}
Typical choices for the utility function include: 1) negative variance of the posterior, 2) Kullback-Leibler (KL) divergence and 3) Wasserstein distance; other utility functions are also possible \cite{Chaloner:1995,McMichael:2022}. Minimizing the variance seeks to minimize the spread about the mean, thereby reducing the uncertainty in the model parameters. Meanwhile, maximizing the KL-divergence and Wasserstein distance  functions seeks controls that maximize the distance between the prior and the posterior distributions, thereby yielding the largest update over the prior distribution. 

\begin{figure}[h]
\centering
\includegraphics[width=0.8\textwidth]{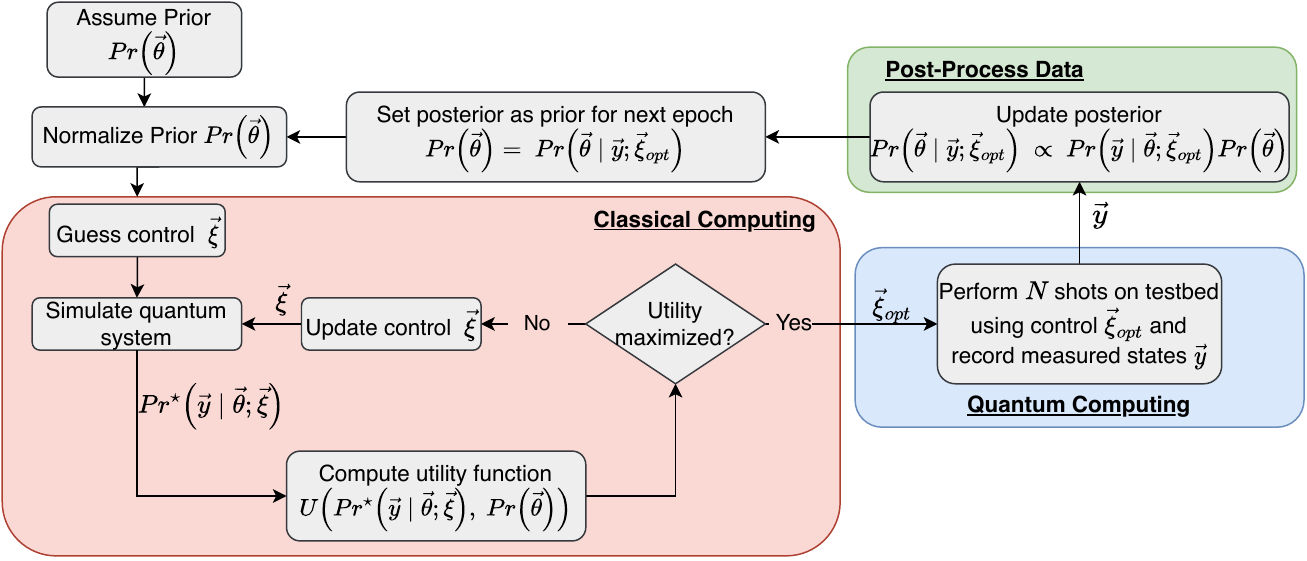}
\captionsetup{singlelinecheck=off,font=footnotesize}
\caption[]{The workflow for our Bayesian experimental design framework}
\label{fig:BExDWorkflow}
\end{figure}

For all such utility functions, the integrals in \eqref{BExD:MarUtil} are often analytically intractable and the \emph{curse of dimensionality} renders standard numerical integration inefficient. Hence, the integral is often approximated using Monte Carlo sampling by drawing $N$ samples $\Vec{\theta}_i \sim \Pr{\Vec{\theta}}$
\eq{BExD:MC}{
    \E{\Vec{\theta}} \approx \sum_{i=1}^N w_i \Vec{\theta}_i 
}
where $w_i$ are the weights and $\sum_i w_i = 1$. These MC samples are also used to construct an approximate discrete representation of the PDF 
\eq{BExD:PDF}{
    \Pr{\Vec{y}|\Vec{\theta}} \approx \sum_{i=1}^N w_i(\Vec{y}) \delta\rbrac{\Vec{\theta} - \Vec{\theta}_i}
}
where $\delta(\cdot)$ is the Dirac delta function. The search for such optimal control $\Vec{\xi}^\star$ requires the solution to an optimization problem \eqref{BExD:UtilOpt} that can be high-dimensional and typically non-convex and multi-modal \Added{\cite{Schoneberger2010}}.
\eq{BExD:UtilOpt}{
    \Vec{\xi}^\star := \arg\max_{\Vec{\xi} \in \Vec{\Xi}} ~~ U(\Vec{\xi})
}
For these reasons, we solve the optimization problem using a non-gradient based differential evolution algorithm. The applicability of the evolutionary method for optimization of utility function in BExD was demonstrated by Hamada \etal \cite{Hamada:2001}. \Added{Schoneberger \etal \cite{Schoneberger2010} demonstrated the robustness of non-gradient based optimization algorithm for optimal experimental design.} An experiment is then performed using the optimal control $\Vec{\xi}^\star$, and the measurements $\Vec{y}$ are used to construct the likelihood and update the posterior distribution at the $n^{th}$ epoch/iteration as
\aligneq{BExD:Update}{
    \Pr{\Vec{\theta} | \Vec{y}_n; \Vec{\xi}_n^\star} & \propto \Pr{\Vec{y}_n | \Vec{\theta}; \Vec{\xi}_n^\star} \Pr{\Vec{\theta} | \Vec{y}_{n-1}; \Vec{\xi}_{n-1}^\star} \\
    & \propto \Added{ \Pr{\Vec{\theta}} \prod_{i=1}^{n} \Pr{\Vec{y}_{i}|\Vec{\theta} ; \Vec{\xi}_{i}^\star}}
}
\Added{To reduce computational cost associated with evaluating the likelihood in regions of low posterior density, the Monte Carlo samples are resampled to redistribute the particles to regions of high posterior density. We make use of the resampling algorithm proposed by Liu and West \cite{Liu2001}}. The expectation of the inferred posterior $\widehat{\Vec{\theta}} :=  \E[{\Pr{\Vec{\theta} | \Vec{y}}}]{\Vec{\theta}}$ yields the unbiased estimator for the model parameters. Figure \ref{fig:BExDWorkflow} presents the complete workflow for Bayesian experimental design.

\subsection{Utility Function: Kullback-Leibler (KL) Divergence}  \label{subsec:BExD:KL}

The Kullback-Leibler (KL) divergence is a measure of distance between two probability distributions and is defined as 
\aligneq{KL:KLD}{
    D_{KL} & = \integral{\Vec{\theta}}{}{\Pr{\Vec{\theta} | \Vec{y}; \Vec{\xi}} \log \rbrac{ \dfrac{\Pr{\Vec{\theta} | \Vec{y}; \Vec{\xi}}}{\Pr{\Vec{\theta}}}} ~d\Vec{\theta}}
}
Note that the KL divergence is not symmetric in the two distributions. At the $n^{th}$ epoch, $\Pr{\Vec{\theta} | \Vec{y}; \Vec{\xi}} \equiv \Pr{\Vec{\theta} | \Vec{y}_n; \Vec{\xi}}$ is the posterior (i.e. \eqref{BExD:Update} for some control vector $\Vec{\xi}$) and $\Pr{\Vec{\theta}} \equiv \Pr{\Vec{\theta} | \Vec{y}_{n-1}; \Vec{\xi^\star_{n-1}}}$ is now the prior (i.e. posterior from previous epoch). Taking the expectation of the KL divergence over all possible outcomes $y$ yields an expression for the \emph{mutual information}
\aligneq{KL:Util}{
    U(\Vec{\xi})  &= \integral{\Vec{y}}{}{ \Pr{\Vec{y} | \Vec{\xi}} \integral{\Vec{\theta}}{}{\Pr{\Vec{\theta} | \Vec{y}; \Vec{\xi}} \log \rbrac{ \dfrac{\Pr{\Vec{\theta} | \Vec{y}; \Vec{\xi}}}{\Pr{\Vec{\theta}}}} ~d\Vec{\theta}} d\Vec{y}}
}
Hence, maximizing the mutual information amounts to identifying control parameters that provide the maximum \emph{information} gain over the prior distribution.

\subsection{Likelihood Function} \label{subsec:BExD:Likelihood}

For clarity of presentation, we denote by $\widetilde{\Vec{\theta}}$ the true distribution of the parameters (i.e. the parameters of the quantum testbed). For a total of $N$ independent shots, let $y_i \sim \Pr{\Vec{y} | \widetilde{\Vec{\theta}}; \Vec{\xi} }$ \Added{be a possible outcome of a shot} (i.e., the state measured for a shot) after applying control $\Vec{\xi}$, $n_i$ the number of times $y_i$ is observed with $\sum_i n_i = N$. Then, the likelihood can be modeled by a multinomial distribution \cite{Granade:2012,McMichael:2022} defined by 
\eq{Likelihood:MNM}{
    \Pr{\Vec{y}|\Vec{\theta}; \Vec{\xi}, N} = \rmatrix{N \\ n_1, n_2, \ldots, n_M} \prod_{i=1}^{N} \Pr{y_i | \Vec{\theta}; \Vec{\xi} }
}
where $\rmatrix{N \\ n_1, n_2, \ldots, n_M} := \dfrac{N!}{\prod_{i=1}^{\abs{\mathbb{S}}} n_i !}$ is the multinomial coefficient, \Added{$y_i \sim\Pr{y_i | \widetilde{\Vec{\theta}}; \Vec{\xi} } = \Pr{y_i | \rho (\widetilde{\Vec{\theta}};\Vec{\xi})} \approx \dfrac{n_i}{N}$ is the probability, estimated from experiments, of observing $y_i$ and $\Pr{y_i | \Vec{\theta}; \Vec{\xi} }$ is the probability of observing $y_i$ computed using \eqref{Quantum:Measurement}} and the numerical model for a given $\Vec{\theta}$ and $\Vec{\xi}$. It should be mentioned that the multinomial coefficient can be omitted since the weights in \eqref{BExD:PDF} are renormalized at each epoch. For a large number of draws $N$, this omission may lead to numerical underflow as the resulting expression would produce near zero probabilities, \Added{particularly when $\Pr{y_i | \Vec{\theta}; \Vec{\xi} }$ is small}. Alternatively, under certain assumptions, the multinomial distribution asymptotically tends to a Gaussian distribution \cite{Gnedenko1998}
\eq{Likelihood:Gau}{
    \Pr{\Vec{y}|\Vec{\theta}; \Vec{\xi}} = \mathcal{N}\rbrac{ \Pr{\Vec{y}|\widetilde{\Vec{\theta}}; \Vec{\xi}} , \Sigma}
}
where $\Sigma$ is the covariance that accounts for errors \Added{between the numerical model and the `true' model (i.e. experiment). This model of the likelihood has been used extensively in Bayesian experimental design \cite{Chaloner:1995, Solonen2012, McMichael:2022} and in Markov Chain Monte Carlo (MCMC) methods. The exact form of $\Sigma$ is not known a priori and must be estimated. Let $\Vec{\mathcal{Y}}(\Vec{\theta}; \Vec{\xi}) = \cbrac{\Pr{y_i|\Vec{\theta}; \Vec{\xi}},i=1,\ldots,N}$ and $\widetilde{\Vec{\mathcal{Y}}}(\Vec{\xi}) = \cbrac{\Pr{y_i|\widetilde{\Vec{\theta}}; \Vec{\xi}},i=1,\ldots,N}$ be vectors of probabilities computed using the numerical model and experimentally measured, respectively, after applying control $\Vec{\xi}$. Then, we estimate $\Sigma$ by means of Monte Carlo sampling over the prior for $\Vec{\theta}$ with $N_{\Vec{\theta}}$ samples and random control $\Vec{\xi}$ with $N_{\Vec{\xi}}$ samples. The sample covariance matrix is then computed as
\eq{Likelihood:Sigma}{
    \Sigma = \dfrac{1}{N_{\Vec{\theta}}} \dfrac{1}{N_{\Vec{\xi}}} \sum_{i=1}^{N_{\Vec{\theta}}} \sum_{j=1}^{N_{\Vec{\xi}}} \rbrac{\Vec{\mathcal{Y}}(\Vec{\theta}_i; \Vec{\xi}_j) - \widetilde{\Vec{\mathcal{Y}}}(\Vec{\xi}_j)} \rbrac{\Vec{\mathcal{Y}}(\Vec{\theta}_i; \Vec{\xi}_j) - \widetilde{\Vec{\mathcal{Y}}}(\Vec{\xi}_j)}^T
}
}

\section{Case Studies for Quantum Characterization} \label{sec:Cases}

\begin{figure}[h]
\centering
\includegraphics[width=0.3\textwidth]{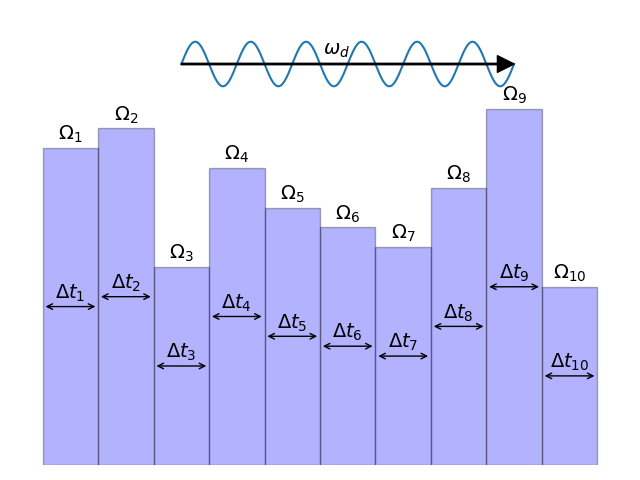}
\captionsetup{singlelinecheck=off,font=footnotesize}
\caption[]{The control pulse parameterization used in this work with $\Omega_j = p_j(t) + i q_j(t)$}
\label{fig:PulseParameterization}
\end{figure}

We now present the problem formulation for verification of our characterization technique using synthetic test problems. We consider an open system modeled by the Lindblad master equation with true value of the parameter set $\Vec{\theta} =\cbrac{\omega,\chi,T_1,T_2}$ to be that of the standard QPU in Quantum Design and Integration Testbed (QuDIT) at LLNL \cite{Martinez2022}. We mainly consider uncertainties in these four parameters, which are the most pronounced noise sources in the experiments, especially for a quantum gate in multi-level system. The QPU has a single transmon in 3D aluminum cavity whose parameters are given \Added{in Tab. \ref{table:QPUParams}}.
\begin{table}[h]
\caption{Parameters of the QPU at LLNL's Quantum Design and Integration Testbed.} \label{table:QPUParams}
\centering
\begin{tabular}{c c c c } 
 \hline
  $\widetilde{\omega}$ (GHz) & $\widetilde{\chi}$ (MHz) & $\widetilde{T_1}$ ($\mu$s)& $\widetilde{T_2}$ ($\mu$s) \\
 \hline 
  4.0108 & 127.8 & 45 & 24 \\
 \hline
\end{tabular}
\end{table}
We consider the cases where the exact model parameters are assumed to be point-estimates or following some distribution. When treating the parameters as point estimates, we take $\widetilde{\Vec{\theta}} = \cbrac{\widetilde{\omega},\widetilde{\chi},\widetilde{T_1},\widetilde{T_2}}$. Each experiment is then performed using this fixed parameter set. In the case where the parameters are assumed to be random variables, we consider their exact distribution to be
\expression{
    \widetilde{\Vec{\theta}} \sim \mathcal{N}(\mu_\Vec{\theta},\Sigma_\Vec{\theta}) := \mathcal{N}(\widetilde{\omega},10^{-6}) \times \mathcal{N}(\widetilde{\chi},10^{-3}) \times \mathcal{N}(\widetilde{T_1},0.2)  \times \mathcal{N}(\widetilde{T_2},0.2)
}
where each experiment is performed using a random sample from the distribution which introduces stochasticity into the model problem. The prior distribution (i.e. distribution at the initialization) is taken to be
\expression{
    \Pr{\Vec{\theta}} = \mathcal{U}_{\omega}(3.5,4.5) \times \mathcal{U}_{\chi}(0.1,0.2) \times \mathcal{U}_{T_1}(30,60) \times \mathcal{U}_{T_2}(20,40)
}
\Added{Since the `true' distribution is not known in practice, the use of an uninformative prior, like the uniform distribution, avoids bias in the inference}. In addition to the \Added{uncertainty} introduced by imposing a distribution over the model parameters, we also introduce shot noise (i.e. sampling uncertainty) by sampling a finite number of outcomes $N$ (in \eqref{Likelihood:MNM}). When omitting shot noise and using exact $\Pr{\Vec{y} | \widetilde{\Vec{\theta}}; \Vec{\xi} }$ (i.e. when $N \rightarrow \infty$), we employ the Gaussian model for the likelihood (\eqref{Likelihood:Gau}). We refer to those experiments performed using point-estimates of parameters and exact measurements (i.e. $N = \infty$) as \Added{\emph{exact}} experiments \Added{and those experiments with uncertainty (either parameter or sampling) as \emph{stochastic} experiments}.
We use a total of 2000 Monte Carlo particles and limit the total number of epochs to 500. \Added{The results with different number of particles is given in Appendix \ref{sec:Appendix}.} It was previously mentioned that the topology of the utility function can be highly multimodal \Added{\cite{Schoneberger2010}}, hence we employ the non-gradient based differential evolution algorithm, since it is easily parallelizable and less susceptable to convergence to local extrema. Each experiment will be defined by a different control pulse that is parameterized as a sequence of piecewise constant segments in the rotating frame, with a drive frequency $\omega_d$ and each segment defined by $p(t)$, $q(t)$ and $\Delta t$ as shown in Fig. \ref{fig:PulseParameterization}. \Added{Although other pulse parameterization, such as spline parameterization \cite{Quandary}, can also be used, the piecewise-constant pulses simplies matrix exponentiation since it can be performed for each segment simultaneously and in parallel. For a finely resolved pulse, this piecewise-constant parameterization will lead to an overwhelming number of parameters that must be optimized, and therefore, a spline-based parameterization is more applicable.} Hence, for an $N_p$ segment pulse, the total number of parameters defining the experiment are $N_T=3N_p+1$. The bounds on the pulse parameters are $p(t), q(t) \in [-12,12]$ MHz, $\Delta t \in [1,30]~\mu$s and $\omega_d \in [3.5,4.5]$. We assume each experiment begins with the qutrit in the ground state $\ket{0}$.
For verification purposes, we compare the converged distributions to the exact, imposed distribution and analyze the error in the model predictions for a reference quantum gate using Swap-02 gate.
\begin{equation}
\mathrm{Swap-02}=
\begin{pmatrix}
0 & 0 & 1\\
0 & 1 & 0\\
1 & 0 & 0
\label{eq:swap02}
\end{pmatrix}
\end{equation}

\begin{figure}[h]
\centering
%
\begin{subfigure}[h]{0.4\textwidth} 
\includegraphics[width=\textwidth]{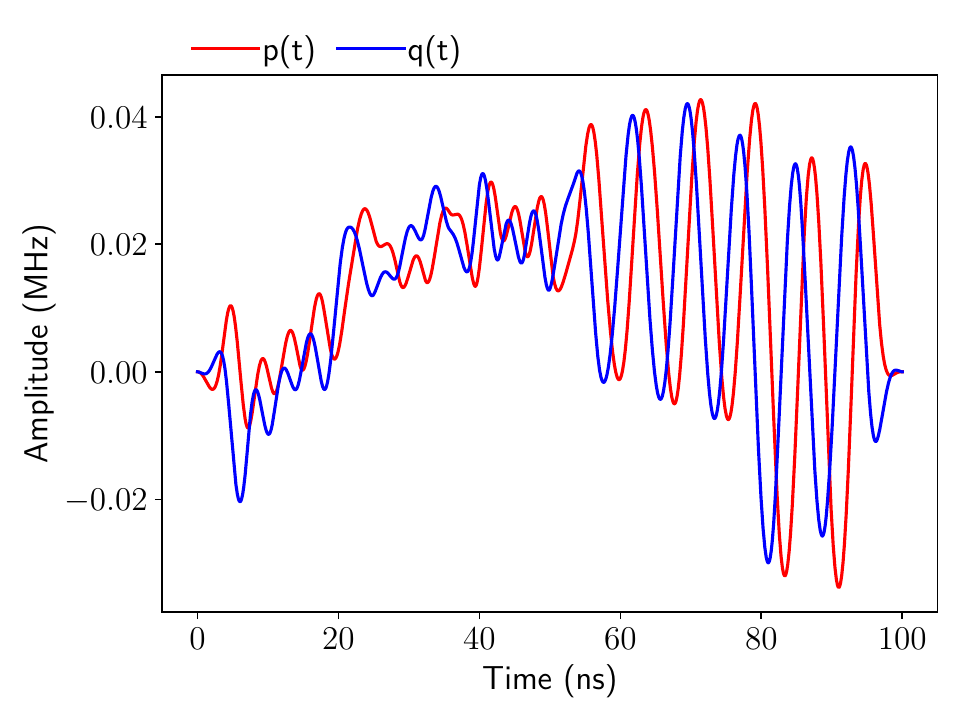} \caption{} \label{fig:Swap02:Control}
\end{subfigure}
\begin{subfigure}[h]{0.4\textwidth} 
\includegraphics[width=\textwidth]{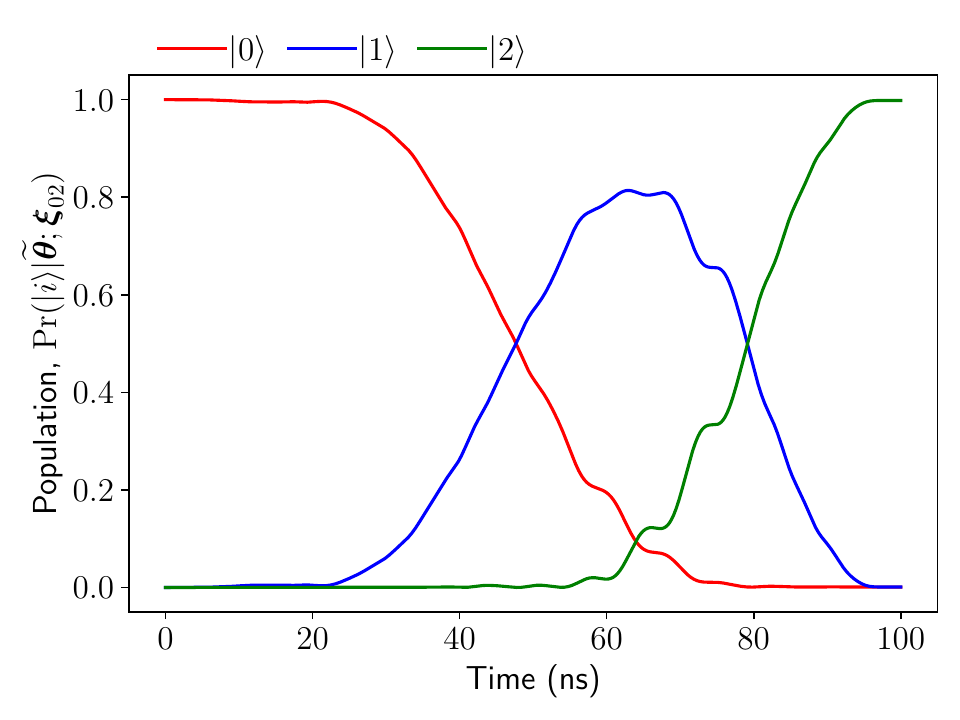} \caption{} \label{fig:Swap02:Population}
\end{subfigure}
\captionsetup{singlelinecheck=off,font=footnotesize}
\caption[]{Time evolution of: (a) controls in rotating frame and (b) population on each state while one Swap02 gate is being played.}
\label{fig:Swap02}
\end{figure}

We define the error in the mean as
\expression{
    \delta(\hat{\mu}) = \E{\widetilde{\mu}} - \hat{\mu}
}
where $\hat{\mu}$ is the unbiased esimator for $\mu \in \Vec{\theta}$. Furthermore, we quantify the errors in the model prediction at each time $t$ as
\expression{
    E(\widetilde{\ket{i}},\ket{i}) = \E[\widetilde{\Vec{\theta}}]{\Pr{\ket{i} | \widetilde{\Vec{\theta}}; \Vec{\xi}_{02} }} - \E[\Vec{\theta}]{\Pr{\ket{i} | \Vec{\theta}; \Vec{\xi}_{02} }}
}
where $\Vec{\xi}_{02}$ is the control pulse for a Swap-02 gate computed using quantum optimal control framework Quandary \cite{Quandary} \Added{with the `true' parameters taken to be point estimates $\widetilde{\Vec{\theta}}$}. The control pulse and the evolution of the population for the Swap-02 gate are shown in Fig. \ref{fig:Swap02}. The short duration of the control pulse (100ns) does not clearly display the dissipative dynamics, such as qubit energy decay over time. For this reason, we repeat the gate 100 times, for a total duration of 10$\mu$s to allow the dissipative dynamics to be more prevalant and amplify the errors.

\section{Results: Characterizing \Added{Qutrit} Systems} \label{sec:QC:Results}

We consider Bayesian experimental design under a fixed control pulse and an approach for \Added{iterative control pulse extension}. In the fixed case, the number of control pulse segments remains constant, whereas in the \Added{iterative pulse extension} case, the number of pulse segments \Added{increase} with epochs.

\subsection{Fixed Pulse Parameterization} \label{subsec:FixedPulse}

\begin{figure}[h]
\centering
%
\begin{subfigure}[h]{0.4\textwidth} 
\includegraphics[width=\textwidth]{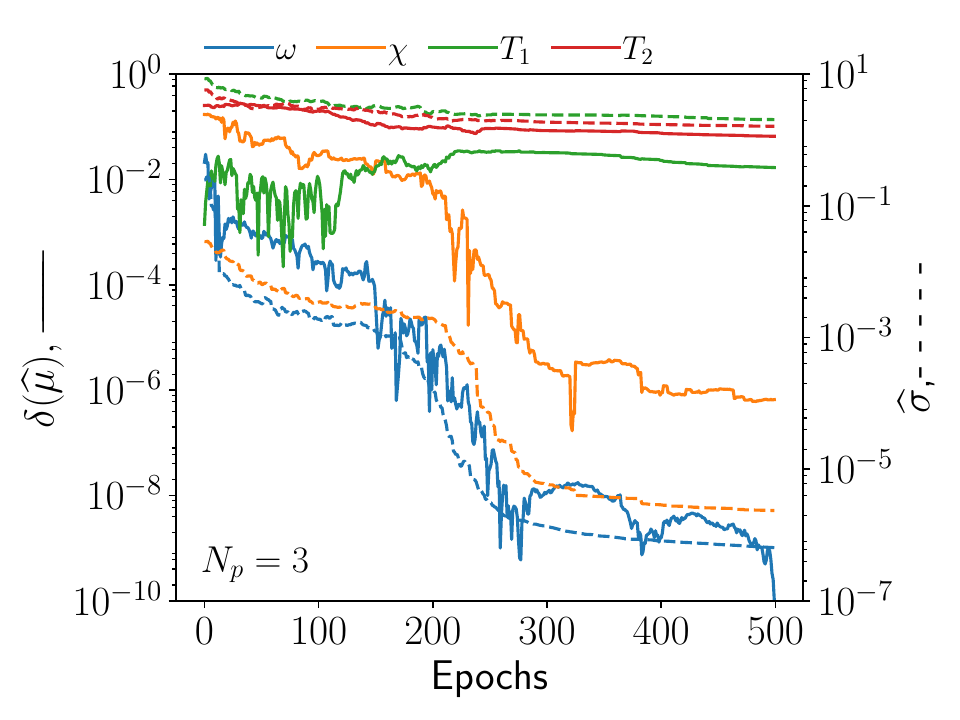} \caption{} \label{fig:QC:Fixed:NoNoise:Conv:3Seg}
\end{subfigure}
\begin{subfigure}[h]{0.4\textwidth} 
\includegraphics[width=\textwidth]{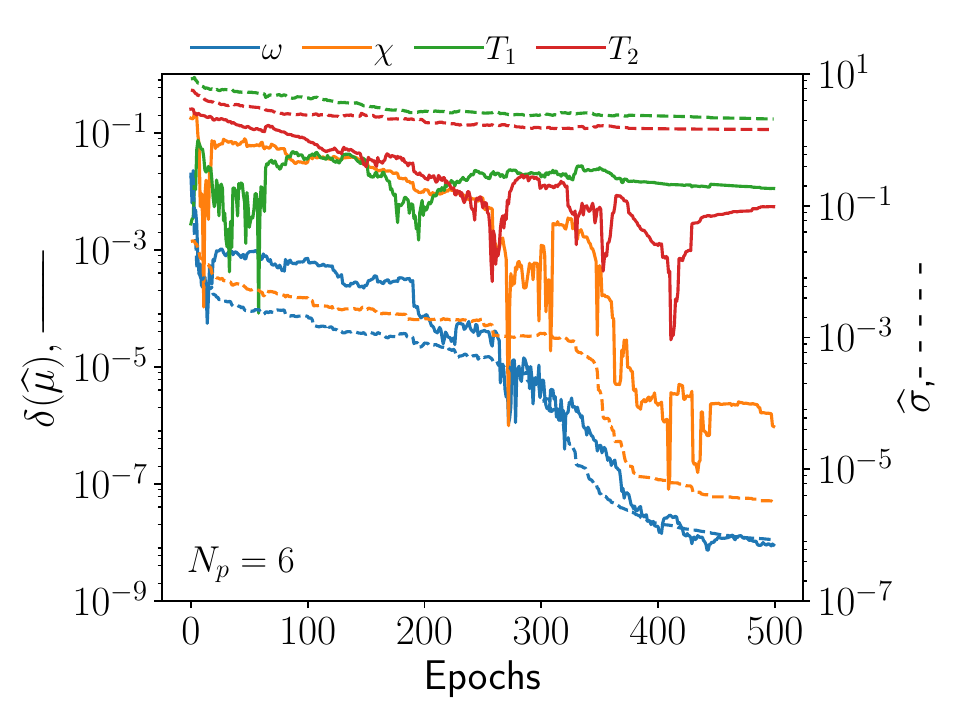} \caption{} \label{fig:QC:Fixed:NoNoise:Conv:6Seg}
\end{subfigure}
\begin{subfigure}[h]{0.4\textwidth} 
\includegraphics[width=\textwidth]{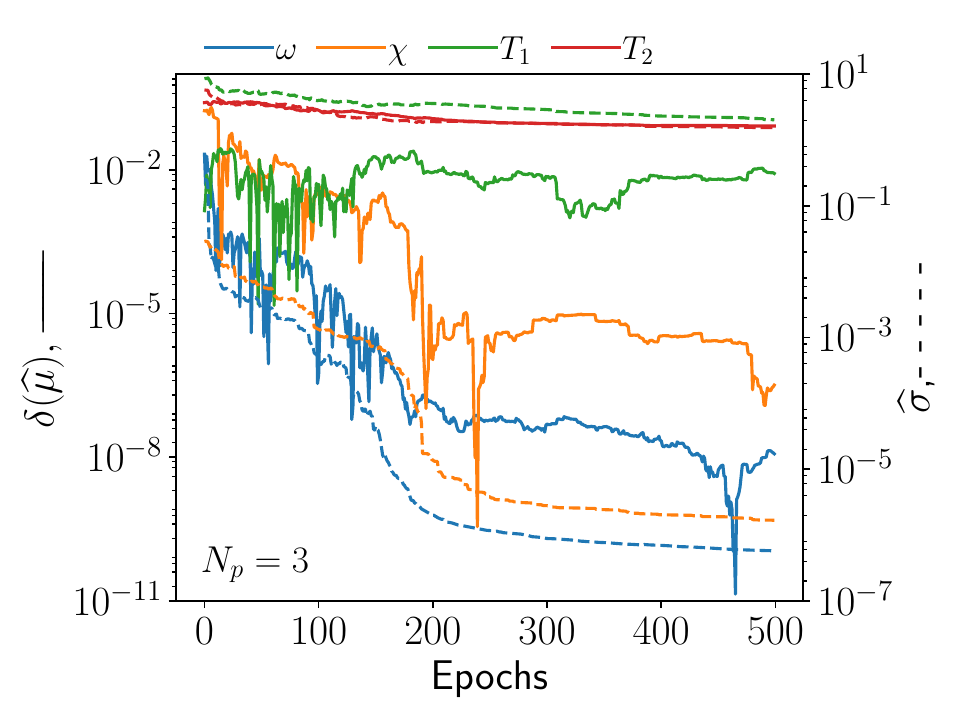} \caption{} \label{fig:QC:Fixed:Noise:Conv:3Seg}
\end{subfigure}
\begin{subfigure}[h]{0.4\textwidth} 
\includegraphics[width=\textwidth]{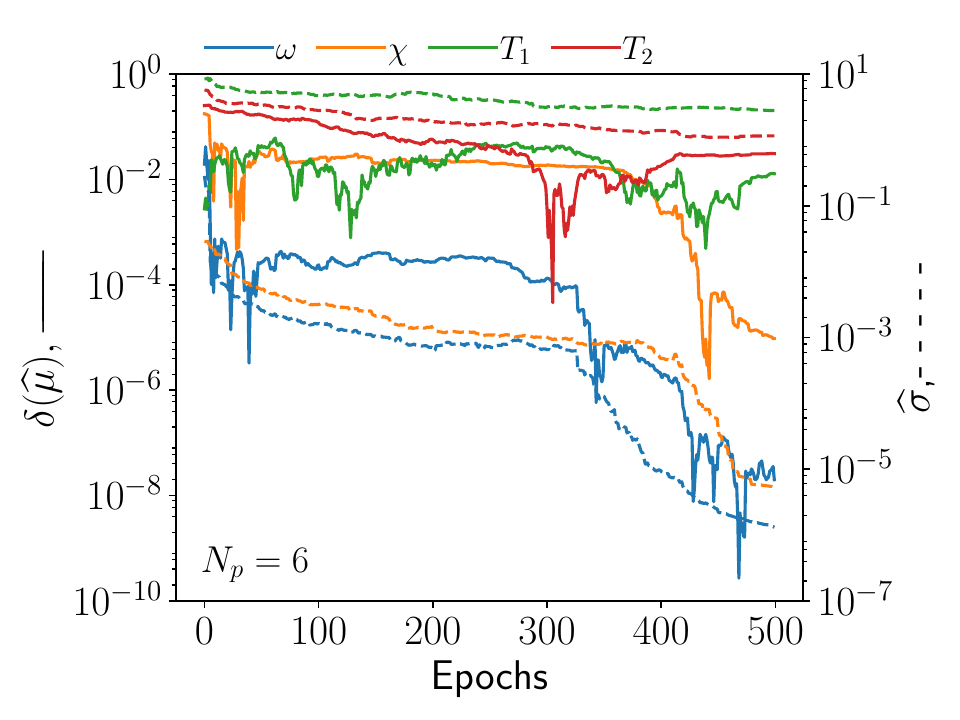} \caption{} \label{fig:QC:Fixed:Noise:Conv:6Seg}
\end{subfigure}
\captionsetup{singlelinecheck=off,font=footnotesize}
\caption[]{Convergence of mean error $\delta(\hat{\mu})$ and standard deviation $\hat{\sigma}$ of the model parameters when performing exact (top) experiments and experiments with parameter \Added{uncertainty} (bottom) using pulses with: (a,c) three segments and (b,d) six segments. \Added{The solid and dashed lines represent the convergence in mean error and standard deviation, respectively.}}
\label{fig:QC:Fixed:Conv}
\end{figure}

\begin{figure}[h]
\centering
%
\begin{subfigure}[h]{0.24\textwidth} 
\includegraphics[width=\textwidth]{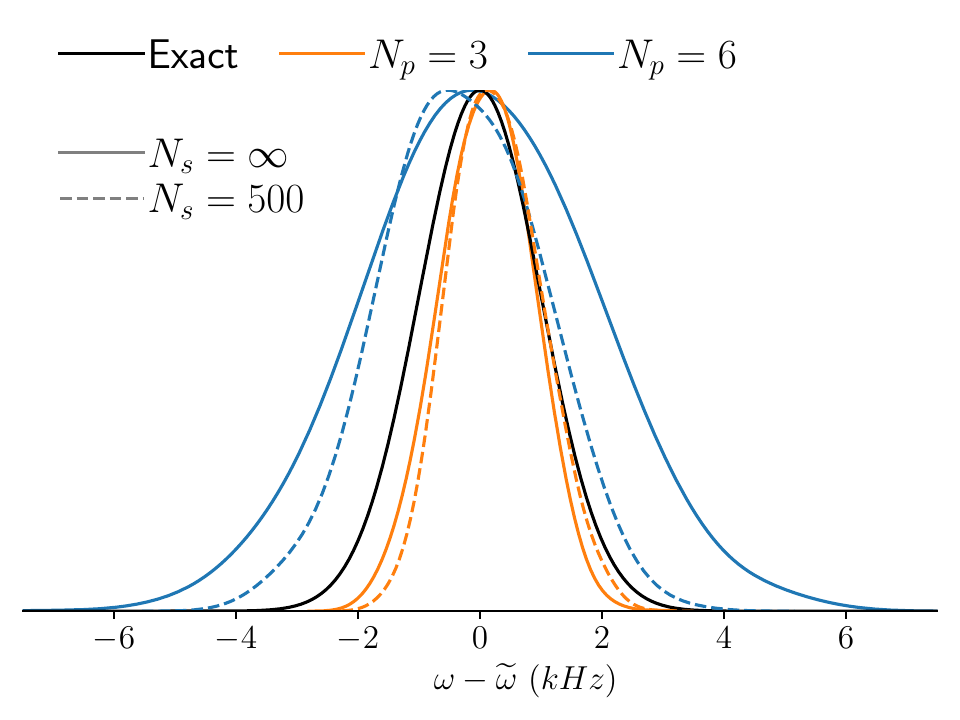} 
\label{fig:QC:Fixed:PDF:omega}
\end{subfigure}
\begin{subfigure}[h]{0.24\textwidth} 
\includegraphics[width=\textwidth]{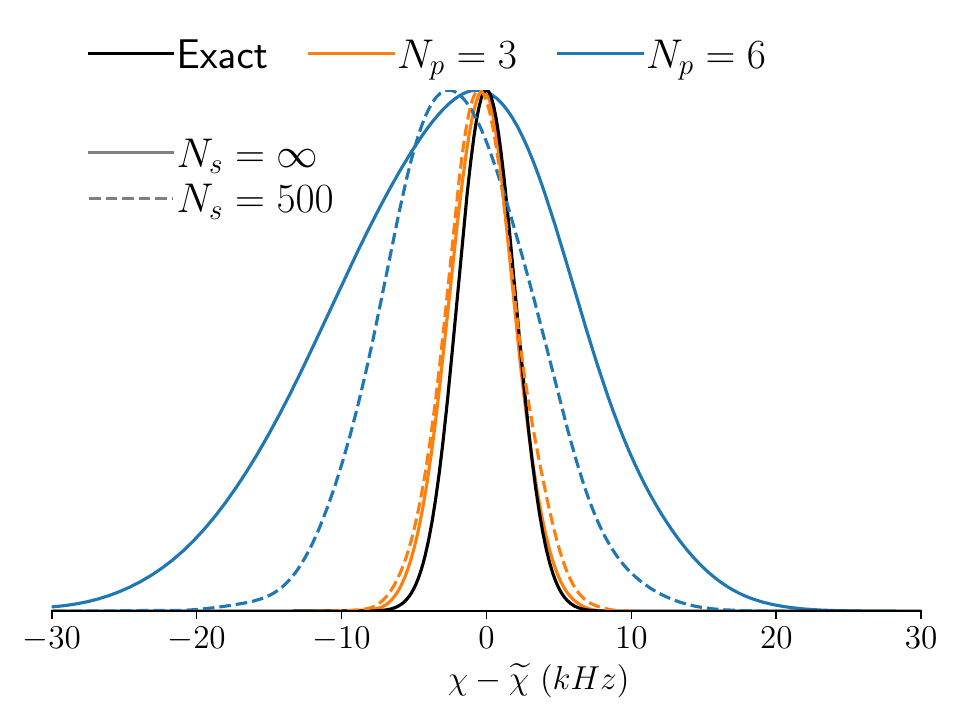} 
\label{fig:QC:Fixed:PDF:chi}
\end{subfigure}
\begin{subfigure}[h]{0.24\textwidth} 
\includegraphics[width=\textwidth]{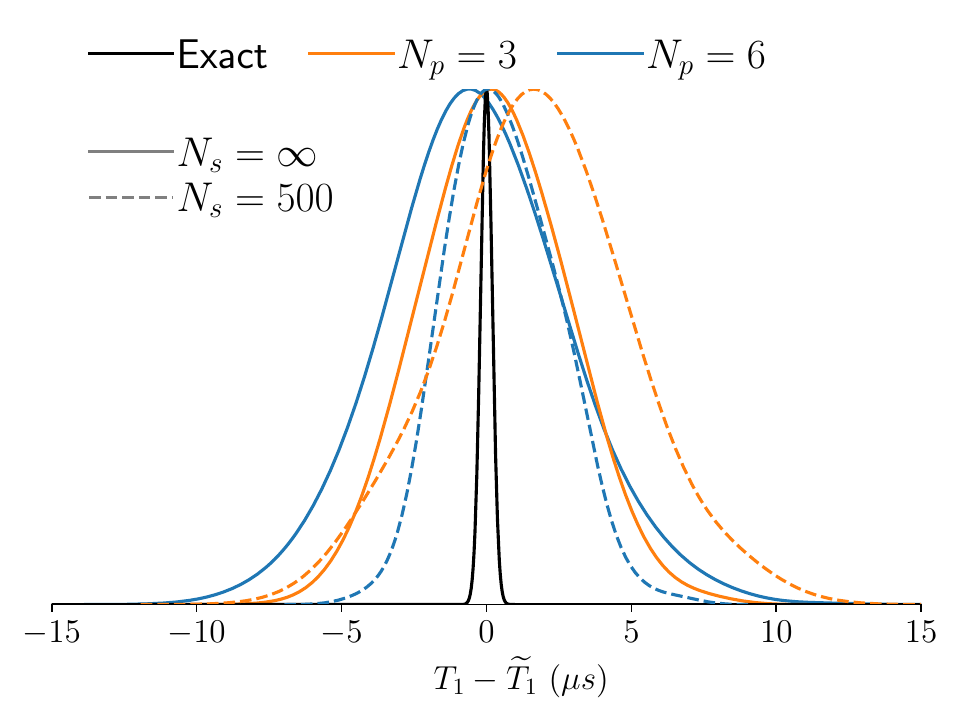} 
\label{fig:QC:Fixed:PDF:T1}
\end{subfigure}
\begin{subfigure}[h]{0.24\textwidth} 
\includegraphics[width=\textwidth]{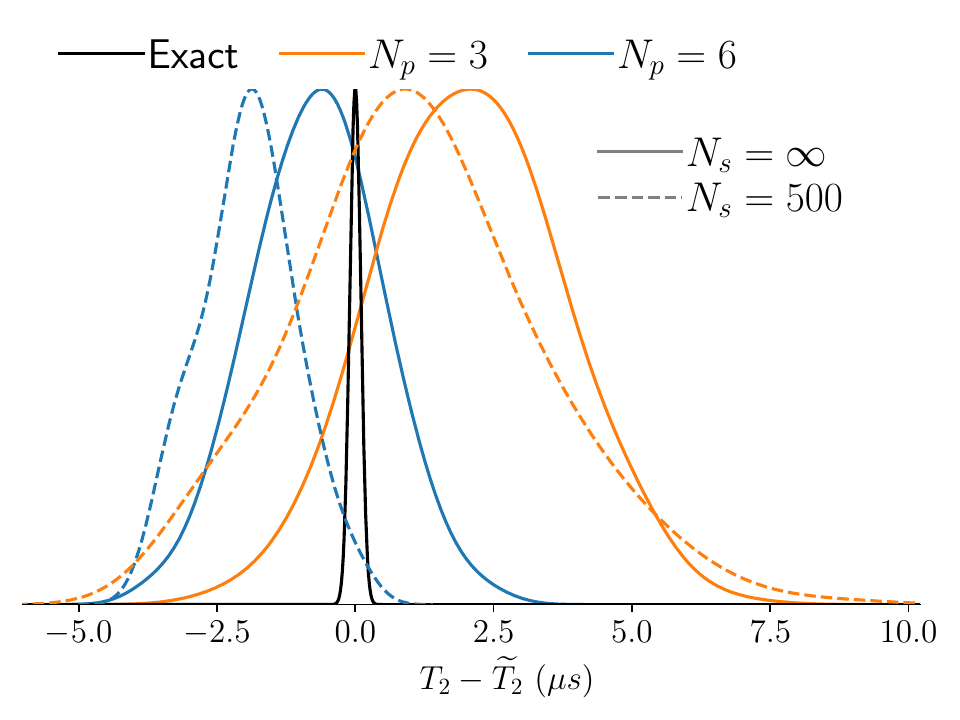} 
\label{fig:QC:Fixed:PDF:T2}
\end{subfigure}
\captionsetup{singlelinecheck=off,font=footnotesize}
\caption[]{Converged parameter distribution when performing \Added{stochastic} experiments using a different number of pulse segments ($N_p$) and different number of shots $N_s$. Note the distributions are translated by the mean of the exact distributions.}
\label{fig:QC:Fixed:PDF}
\end{figure}

We first investigate the effectiveness of our approach under a fixed pulse parameterization and the problem formulation presented in Sec. \ref{sec:Cases}. Figure \ref{fig:QC:Fixed:Conv} shows the mean error and the standard deviation of the estimated distributions using different number of control pulse segments under different sources of uncertainty. Immediately we see that the mean error and the standard deviation for $\omega$ and $\chi$ distributions are orders of magnitude smaller than those for the distributions of $T_1$ and $T_2$. Furthermore, variances of all four parameter distributions decrease near-monotonically. In \Added{both the exact and stochastic experiments}, the transition frequency and anharmonicity show a faster rate of convergence than the decoherence parameters, with several orders of magnitude reduction in only the first few epochs. We see a better overall convergence in $T_1$ and $T_2$ when using a more robust pulse parameterization with a larger number of segments, but with a slight decrease in the accuracy of $\omega$ and $\chi$ estimates. Control pulses with a greater number of segments lead to longer pulse durations and better expose the dissipative dynamics that are present on larger timescales. This suggests that an adaptive pulse parameterization - where short pulse sequences are used to accurately estimate $\omega$ and $\chi$, then longer sequences are used to improve the $T_1$ and $T_2$ estimates - may be beneficial.
Figure \ref{fig:QC:Fixed:PDF} shows the exact (imposed) and the converged distribution for the four parameters using a different number of pulse segments $N_p$ and shots $N_s$. We see a good agreement in the distributions for $\omega$ and $\chi$ whereas the distributions of $T_1$ and $T_2$ feature larger variances. The mean of $T_1$ has better agreement with that of the exact distribution, whereas the distribution of $T_2$ displays both a larger error in the mean and a wider spread. As expected, the \Added{uncertainty} introduced by sampling a finite number of shots introduces additional errors; this often manifests as a larger shift in mean or larger variances. Furthermore, we see that shorter pulse sequences yield better estimates for $\omega$ and $\chi$ whereas longer pulse sequences result in better estimates for $T_1$ and $T_2$. The statistical moments of the converged distribution reported in Tab. \ref{table:Fixed:Parameters} show that the frequency parameters are less susceptable to shot noise than the decoherence times. 

\begin{table}[h]
\caption{Mean and standard deviation (in the brackets) of the parameters estimated using a fixed pulse parameterization} \label{table:Fixed:Parameters}
\centering
\begin{tabular}{c | c | c | c | c | c} 
\hline
\Added{Uncertainty} Type & Segments & $\omega$ (GHz) & $\chi$ (MHz) & $T_1$ ($\mu$s) & $T_2$ ($\mu$s) \\
\hline
\multirow{2}{*}{None} & 3 & 4.0108 (8.950\Added{$\times 10^{-7}$}) & 127.8 (3.672\Added{$\times 10^{-6}$}) & 43.806 (2.397) & 26.011 (1.789) \\
                      & 6 & 4.0108 (1.192\Added{$\times 10^{-6}$}) & 127.8 (5.458\Added{$\times 10^{-6}$}) & 45.534 (2.255) & 23.936 (1.472) \\
\hline
\multirow{2}{*}{Parameter} & 3 & 4.0108 (7.169\Added{$\times 10^{-7}$}) & 127.8 (2.100\Added{$\times 10^{-6}$}) & 45.289 (2.387) & 26.052 (1.624) \\
                           & 6 & 4.0108 (1.801\Added{$\times 10^{-6}$}) & 127.8 (7.796\Added{$\times 10^{-6}$}) & 44.677 (3) & 23.308 (1.142) \\
\hline
\multirow{2}{*}{Parameter and Shot} & 3 & 4.0108 (7.065\Added{$\times 10^{-7}$}) & 127.8 (2.335\Added{$\times 10^{-6}$}) & 46.738 (3.081) & 25.123 (2.169) \\
                                    & 6 & 4.0108 (1.234\Added{$\times 10^{-6}$}) & 127.8 (4.794\Added{$\times 10^{-6}$}) & 45.584 (1.805) & 22.063 (0.8545) \\
\hline
\end{tabular}
\end{table}

\begin{figure}[h]
\centering
%
\begin{subfigure}[h]{0.32\textwidth} 
\includegraphics[width=\textwidth]{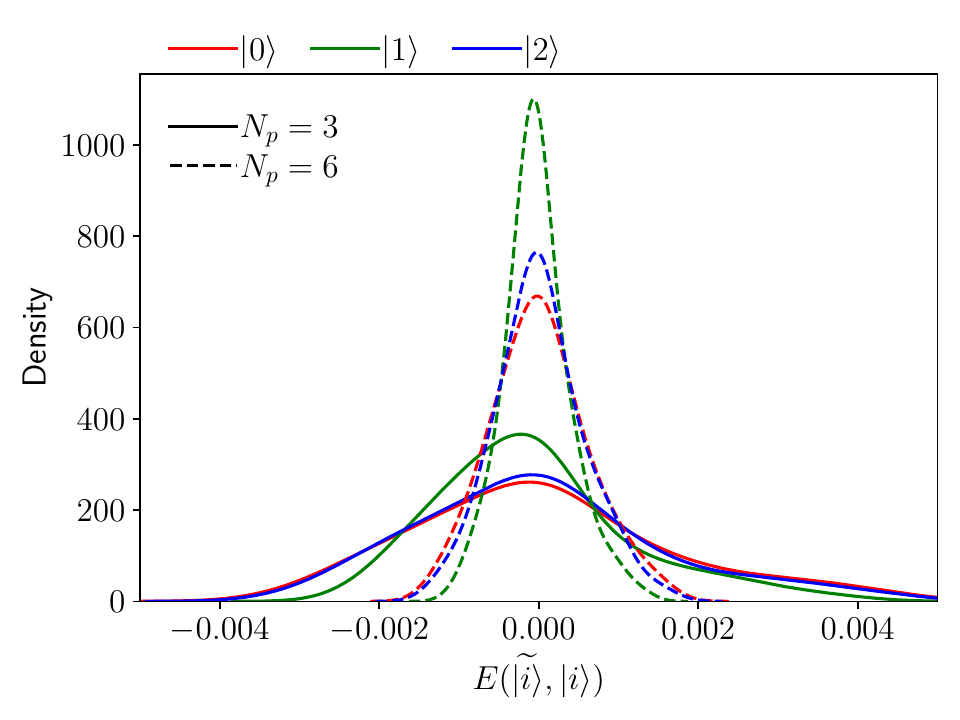} \caption{} \label{fig:QC:Fixed:NoNoise:Error}
\end{subfigure}
\begin{subfigure}[h]{0.32\textwidth} 
\includegraphics[width=\textwidth]{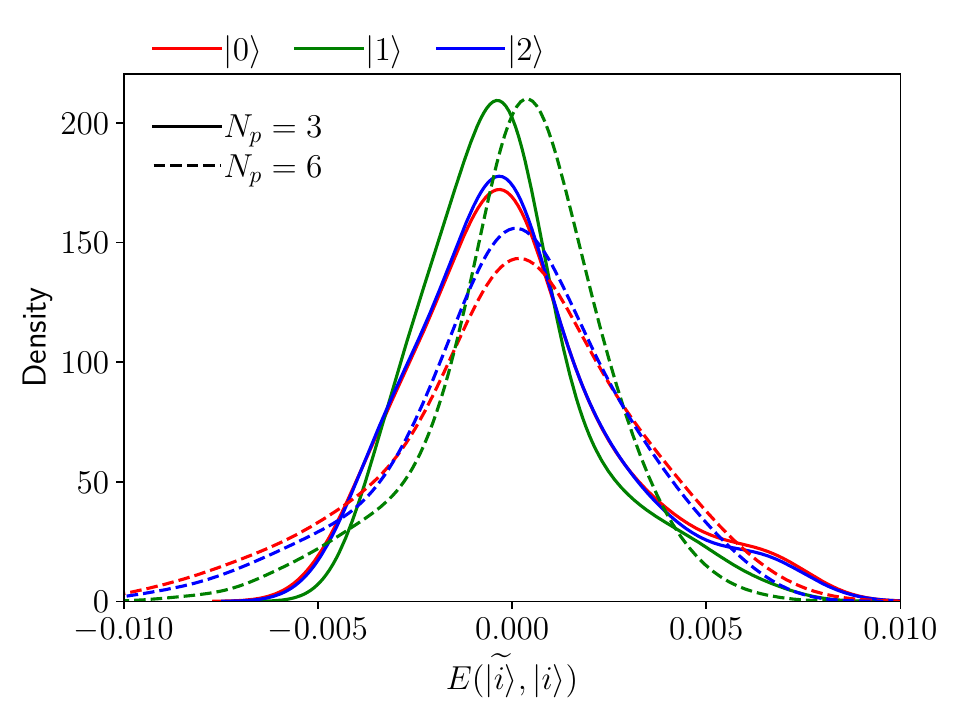} \caption{} \label{fig:QC:Fixed:PNoise:Error}
\end{subfigure}
\begin{subfigure}[h]{0.32\textwidth} 
\includegraphics[width=\textwidth]{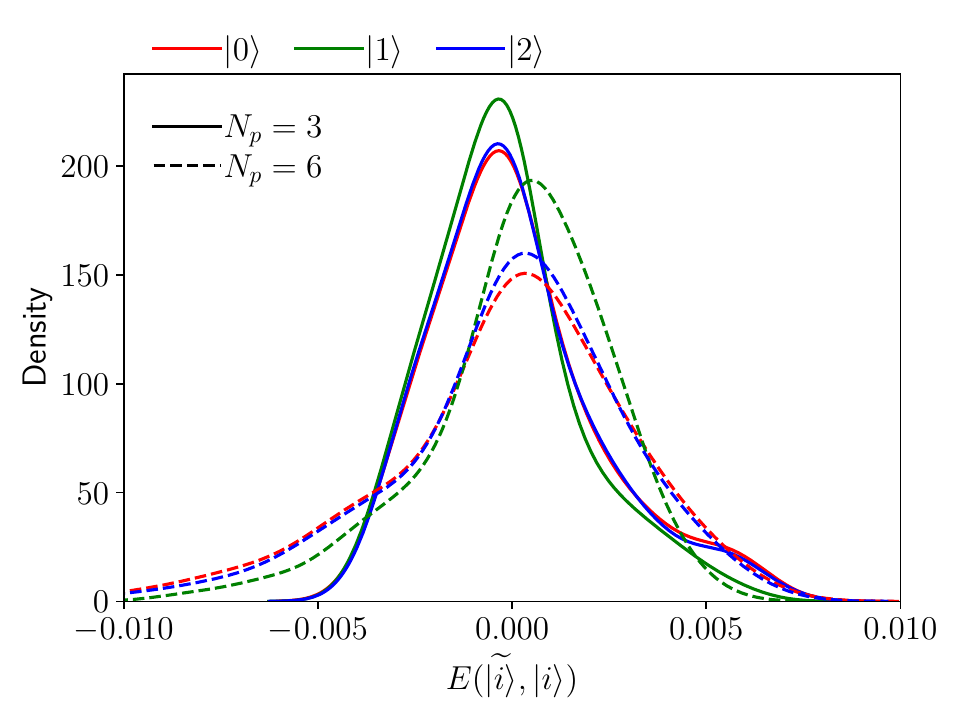} \caption{} \label{fig:QC:Fixed:PSNoise:Error}
\end{subfigure}
\captionsetup{singlelinecheck=off,font=footnotesize}
\caption[]{Distribution of the error in the population when using a different number of pulse segments and performing: a) \Added{exact} experiments, b) experiments with parameter \Added{uncertainty}, and c) experiments with parameter \Added{uncertainty} and shot noise.}
\label{fig:QC:Fixed:Error:PDF}
\end{figure}

When model parameters are not unique, quantifying the accuracy using only errors in the parameters is not sufficient. In such cases, it is imperative to quantify the accuracy of model response/predictions. Figure \ref{fig:QC:Fixed:Error:PDF} shows the probability density of the errors in the population due to \Added{different sources of uncertainty}, pulse segments, and number of shots. Furthermore, the mean and standard deviation of the errors is also presented in Tab. \ref{table:Fixed:Error}. In the case of \Added{exact} experiments, we see that longer pulse sequences yield model parameters that result in more accurate predictions (i.e. higher densities for smaller errors). This is due to the more accurate estimation of $T_1$ and $T_2$ and therefore, more accurate representation of the dissipative dynamics that are present in longer duration pulses (i.e., larger gate repitition). When performing estimation using \Added{stochastic} experiments, we see that the error distribution for all population and pulse parameterization are similar, and are an order of magnitude larger than that of \Added{exact} predictions. Furthermore, despite the lack of accuracy in the decoherence parameters compared to the frequency parameters, the \Added{mathematical} model with estimated parameters yields sufficiently accurate predictions with mean errors on the order of $10^{-4}$ and variances on the order of $10^{-3}$.

\begin{table}[h]
\caption{Mean and standard deviation (in the brackets) of the error using a fixed pulse parameterization} \label{table:Fixed:Error}
\centering
\begin{tabular}{c | c | c | c | c} 
\hline
\Added{Uncertainty} Type & Segments & $\ket{0}$ & $\ket{1}$ & $\ket{2}$\\
\hline
\multirow{2}{*}{None} & 3 & 5.31\Added{$\times 10^{-5}$} (1.60\Added{$\times 10^{-3}$}) & -6.39\Added{$\times 10^{-5}$} (1.20\Added{$\times 10^{-3}$}) & 1.08\Added{$\times 10^{-5}$} (1.53\Added{$\times 10^{-3}$}) \\
                          & 6 & 8.60\Added{$\times 10^{-6}$} (5.93\Added{$\times 10^{-4}$}) & -1.98\Added{$\times 10^{-5}$} (4.08\Added{$\times 10^{-4}$}) & 1.12\Added{$\times 10^{-5}$} (5.47\Added{$\times 10^{-4}$}) \\
\hline
\multirow{2}{*}{Parameter} & 3 & 1.18\Added{$\times 10^{-4}$} (2.56\Added{$\times 10^{-3}$}) & -1.87\Added{$\times 10^{-4}$} (2.10\Added{$\times 10^{-3}$}) & 6.85\Added{$\times 10^{-5}$} (2.49\Added{$\times 10^{-3}$}) \\
                           & 6 & -6.36\Added{$\times 10^{-5}$} (3.03\Added{$\times 10^{-3}$}) & 1.09\Added{$\times 10^{-4}$} (2.21\Added{$\times 10^{-3}$}) & -4.58\Added{$\times 10^{-5}$} (2.78\Added{$\times 10^{-3}$}) \\
\hline
\multirow{2}{*}{Parameter and Shot} & 3 & 1.17\Added{$\times 10^{-4}$} (2.15\Added{$\times 10^{-3}$}) & -2.03\Added{$\times 10^{-4}$} (1.88\Added{$\times 10^{-3}$}) & 8.58\Added{$\times 10^{-5}$} (2.11\Added{$\times 10^{-3}$}) \\
                                    & 6 & -1.07\Added{$\times 10^{-4}$} (2.99\Added{$\times 10^{-3}$}) & 1.61\Added{$\times 10^{-4}$} (2.31\Added{$\times 10^{-3}$}) & -5.40\Added{$\times 10^{-5}$} (2.86\Added{$\times 10^{-3}$}) \\
\hline
\end{tabular}
\end{table}

\subsection{\Added{Iterative Pulse Extension}} \label{subsec:AdaptivePulse}
\begin{figure}[h]
\centering
%
\begin{subfigure}[h]{0.32\textwidth} 
\includegraphics[width=\textwidth]{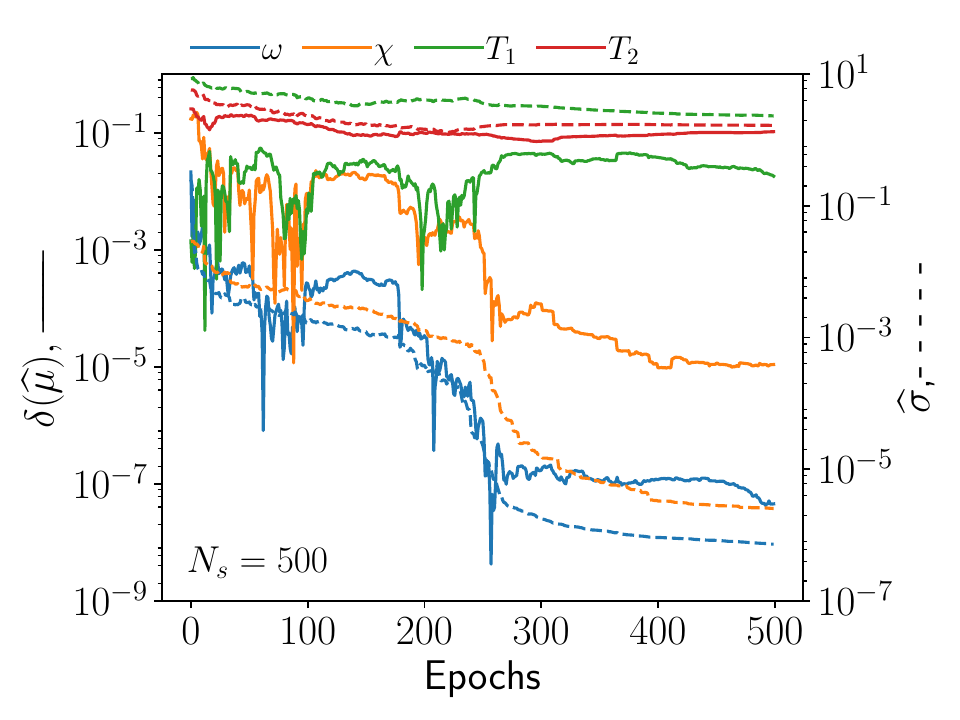} 
\caption{} \label{fig:QC:Adapt:Noise:Conv:500Shots}
\end{subfigure}
\begin{subfigure}[h]{0.32\textwidth} 
\includegraphics[width=\textwidth]{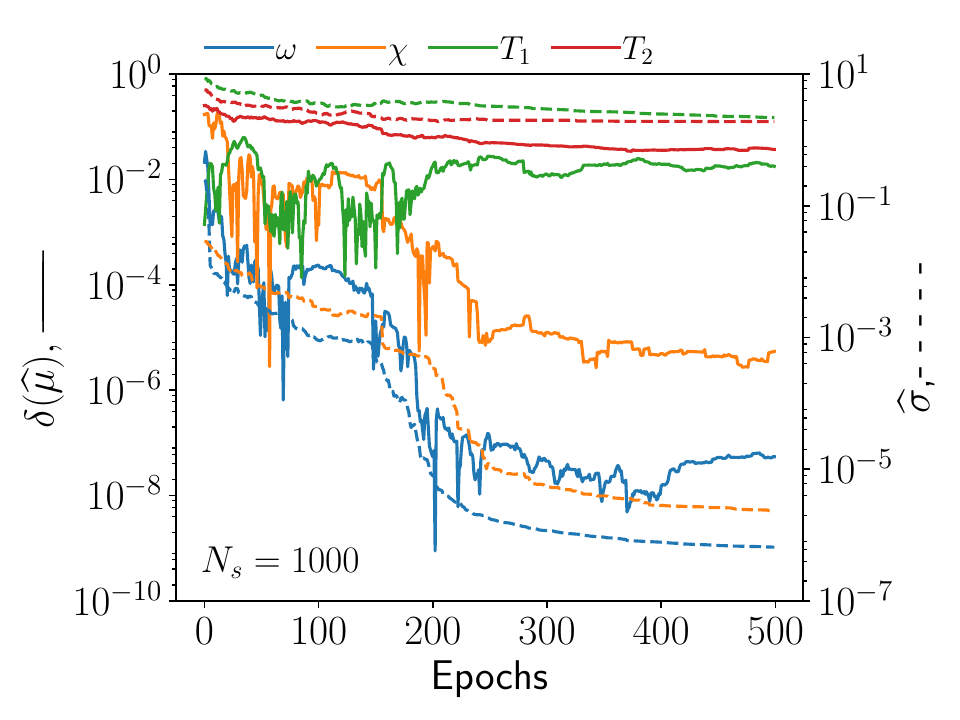} 
\caption{} \label{fig:QC:Adapt:Noise:Conv:1000Shots}
\end{subfigure}
\begin{subfigure}[h]{0.32\textwidth} 
\includegraphics[width=\textwidth]{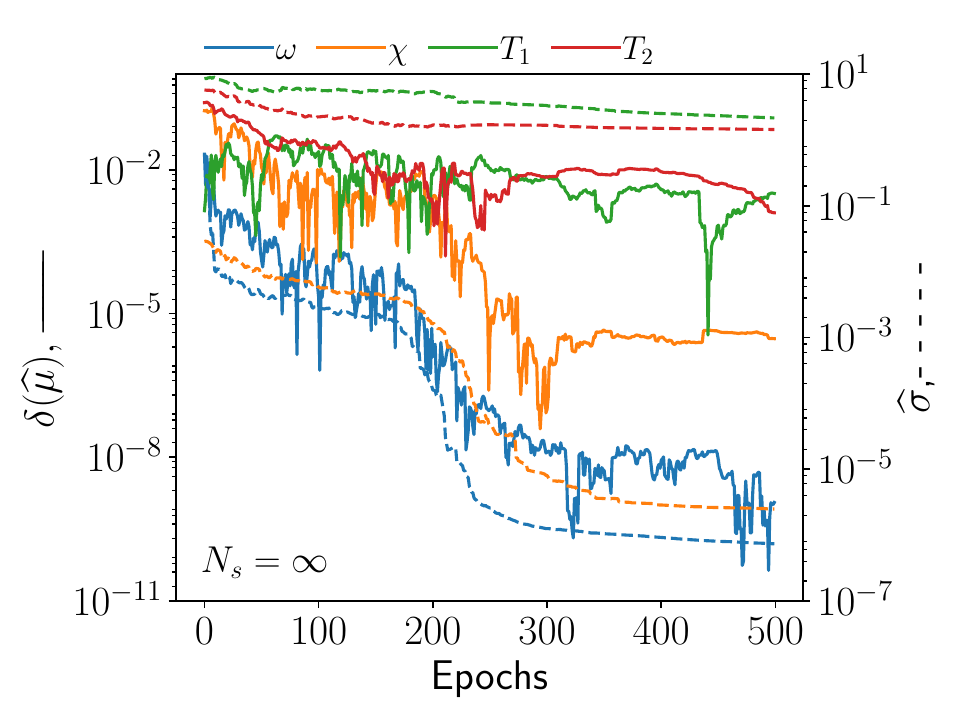} 
\caption{} \label{fig:QC:Adapt:Noise:Conv:0Shots}
\end{subfigure}
\captionsetup{singlelinecheck=off,font=footnotesize}
\caption[]{Convergence of mean error $\delta(\hat{\mu})$ and standard deviation $\hat{\sigma}$ of the model parameters when performing experiments with parameter \Added{uncertainty} using : (a) 500 shots, (b) 1000 shots and (c) $\infty$ shots. \Added{The solid and dashed lines represent the convergence in mean error and standard deviation, respectively.}}
\label{fig:QC:Adaptive:Conv}
\end{figure}

Since different pulse parameterization greatly affects the accuracy of the estimated parmeter distributions asymmetrically, we now investigate \Added{iterative pulse extensions} to improve convergence in all four parameters. We begin with three segments and increase the number of segments by one every 100 epochs. We only consider cases with parameter (defined as random variables) and shot noise, with 500, 1000, and $\infty$ shots. 
Figure \ref{fig:QC:Adaptive:Conv} shows the convergence history in the mean and the standard deviation of the four parameters. Although the accuracy of the frequencies is comparable to the fixed pulse parameterization, the decoherence parameters are better estimated with \Added{iterative pulse extensions}. The mean error in the decoherence parameters are an order of magnitude lower than the fixed pulse parameterization. When we introduce shot noise, we see the accuracy is comparable to the fixed pulse parameterization. Table \ref{table:Adaptive:Parameters} shows the mean and standard deviation of the four parameters. The improved accuracy of $T_1$ and $T_2$ estimation is clear using a moderate number of shots. As expected, large sampling errors (from severe undersampling) can greatly affect the accuracy of the parameter estimates.
\begin{table}[h]
\caption{Mean and standard deviation (in the brackets) of the parameters using \Added{iterative pulse extensions.}} \label{table:Adaptive:Parameters}
\centering
\begin{tabular}{c | c | c | c | c} 
\hline
Shots & $\omega$ (GHz) & $\chi$ (MHz) & $T_1$ ($\mu$s) & $T_2$ ($\mu$s)\\
\hline
500 & 4.0108 (9.260\Added{$\times 10^{-7}$}) & 127.8 (3.645\Added{$\times 10^{-6}$}) & 43.242 (2.592) & 26.196 (1.719) \\
1000 & 4.0108 (6.813\Added{$\times 10^{-7}$}) & 127.8 (2.469\Added{$\times 10^{-6}$}) & 44.185 (2.304) & 24.904 (1.928) \\
$\infty$ & 4.0108 (8.324\Added{$\times 10^{-7}$}) & 127.8 (2.686\Added{$\times 10^{-6}$}) & 44.951 (2.416) & 24.171 (1.494) \\
\hline
\end{tabular}
\end{table}

\begin{figure}[h]
\centering
%
\begin{subfigure}[h]{0.24\textwidth} 
\includegraphics[width=\textwidth]{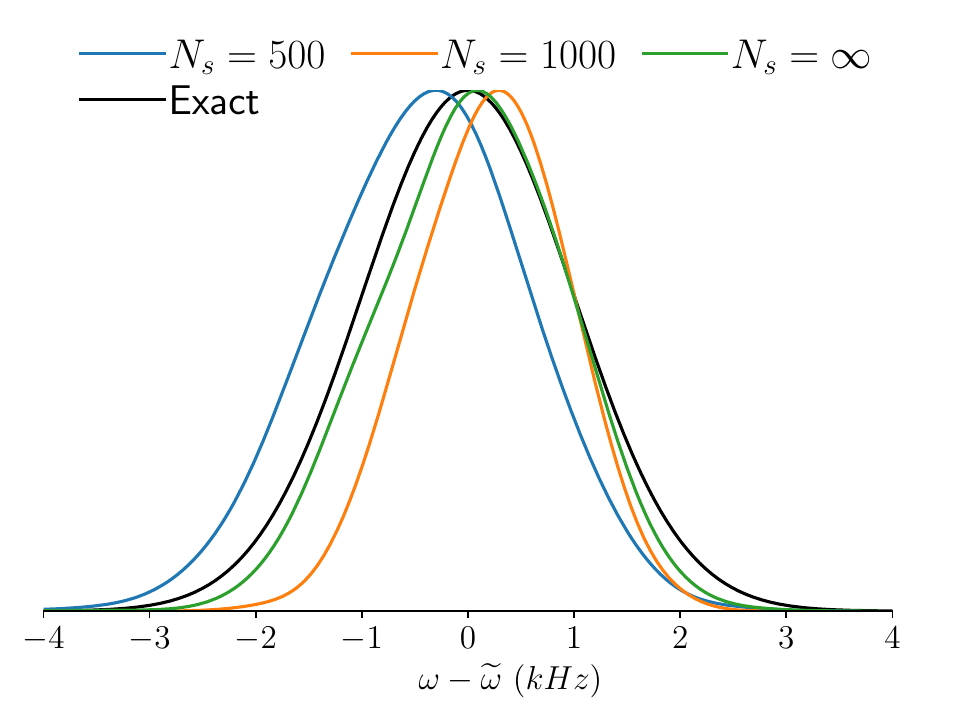} 
\end{subfigure}
\begin{subfigure}[h]{0.24\textwidth} 
\includegraphics[width=\textwidth]{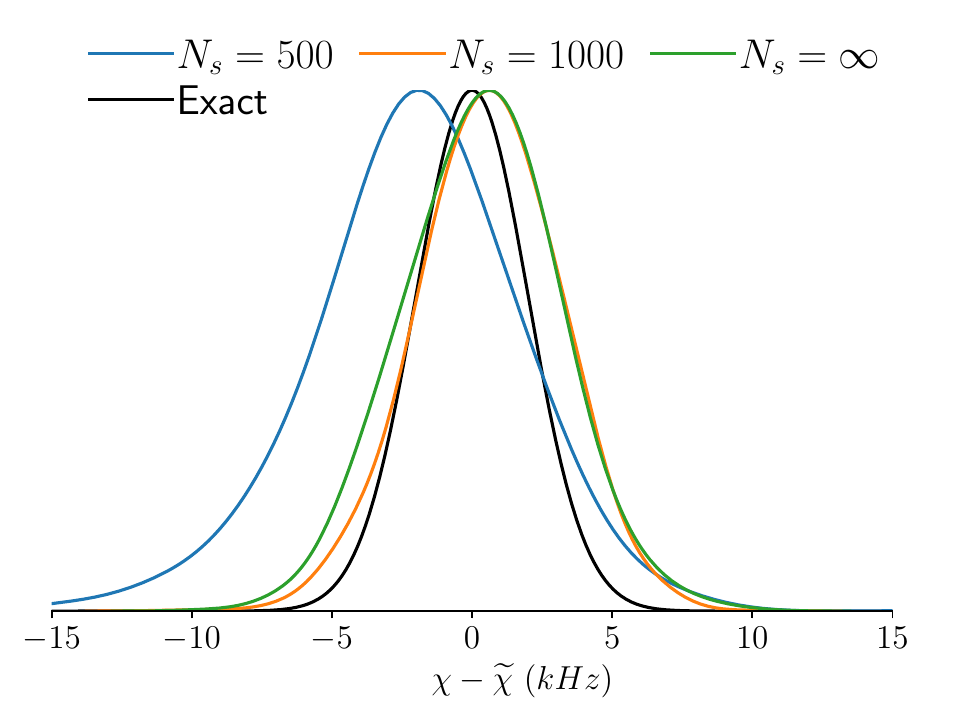} 
\end{subfigure}
\begin{subfigure}[h]{0.24\textwidth} 
\includegraphics[width=\textwidth]{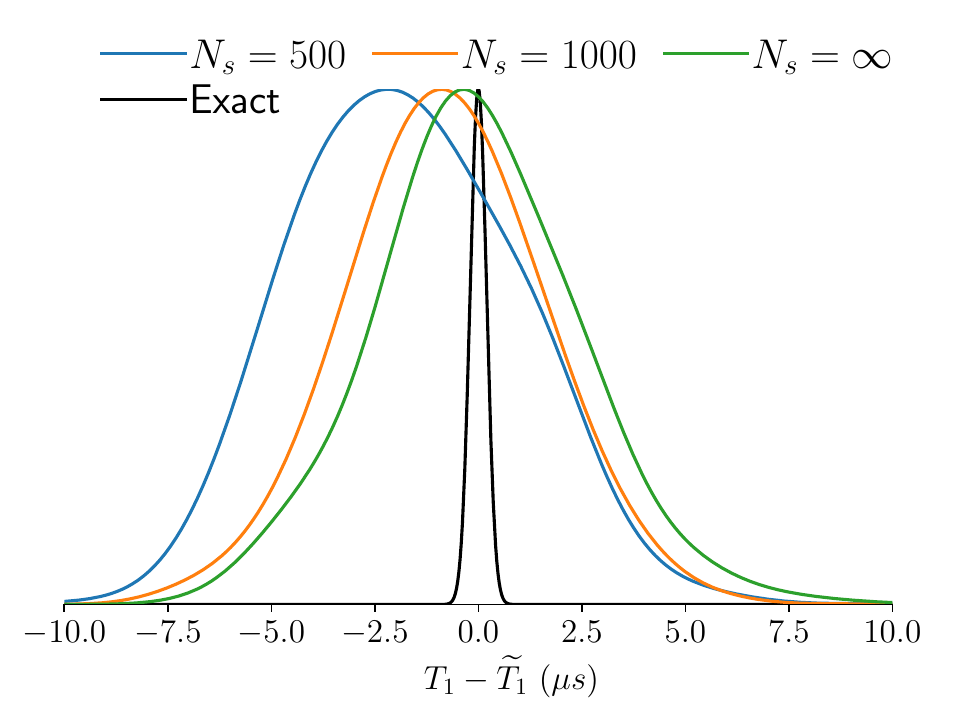} 
\end{subfigure}
\begin{subfigure}[h]{0.24\textwidth} 
\includegraphics[width=\textwidth]{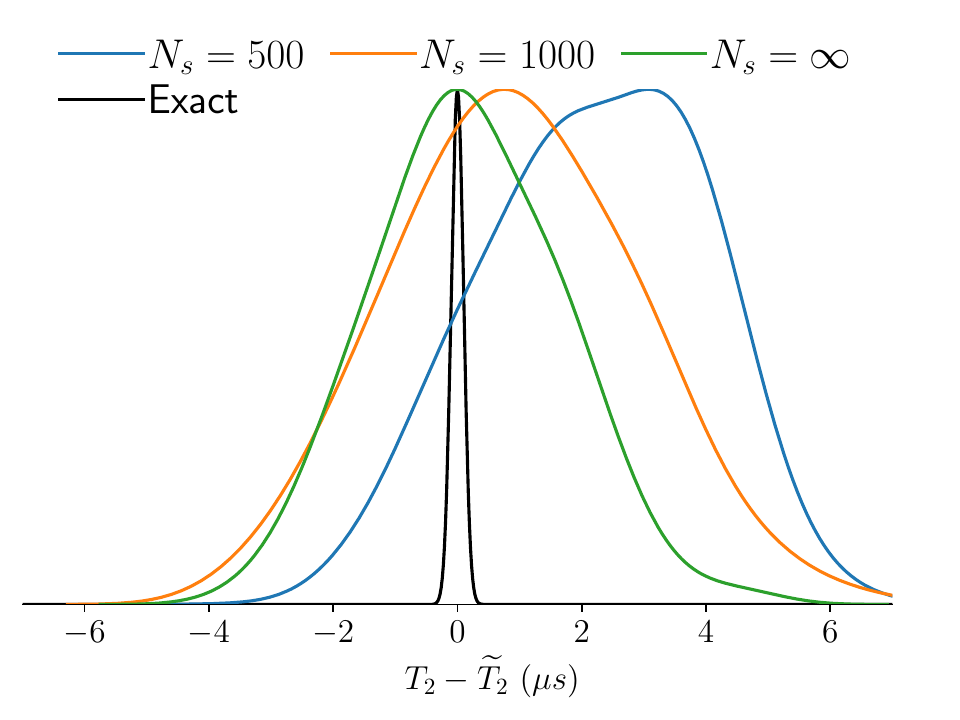} 
\end{subfigure}
\begin{subfigure}[h]{0.32\textwidth} 
\includegraphics[width=\textwidth]{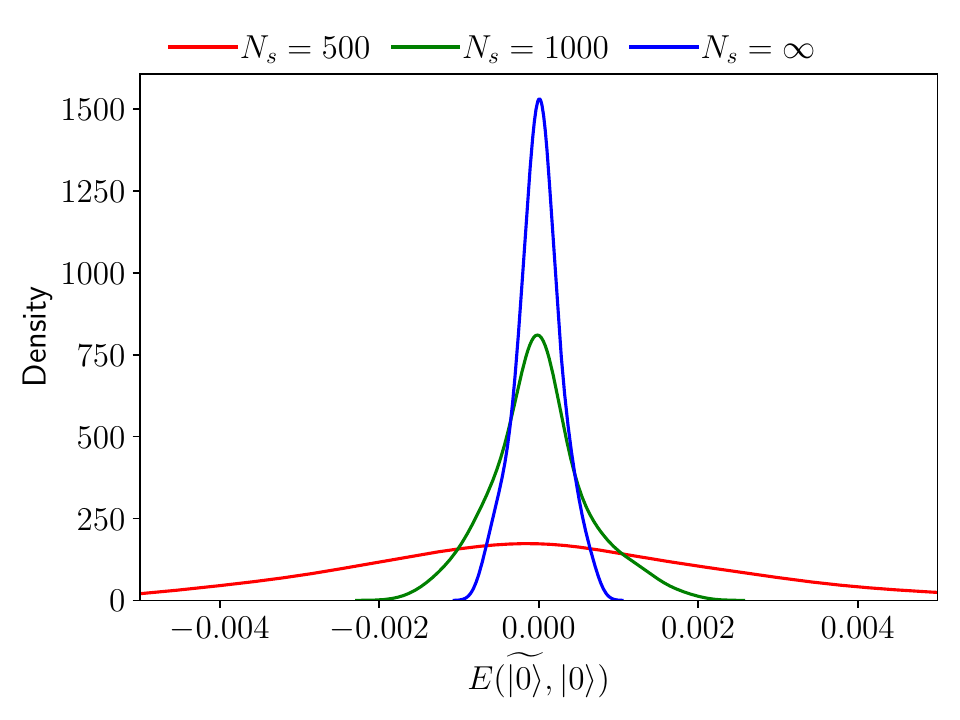} 
\end{subfigure}
\begin{subfigure}[h]{0.32\textwidth} 
\includegraphics[width=\textwidth]{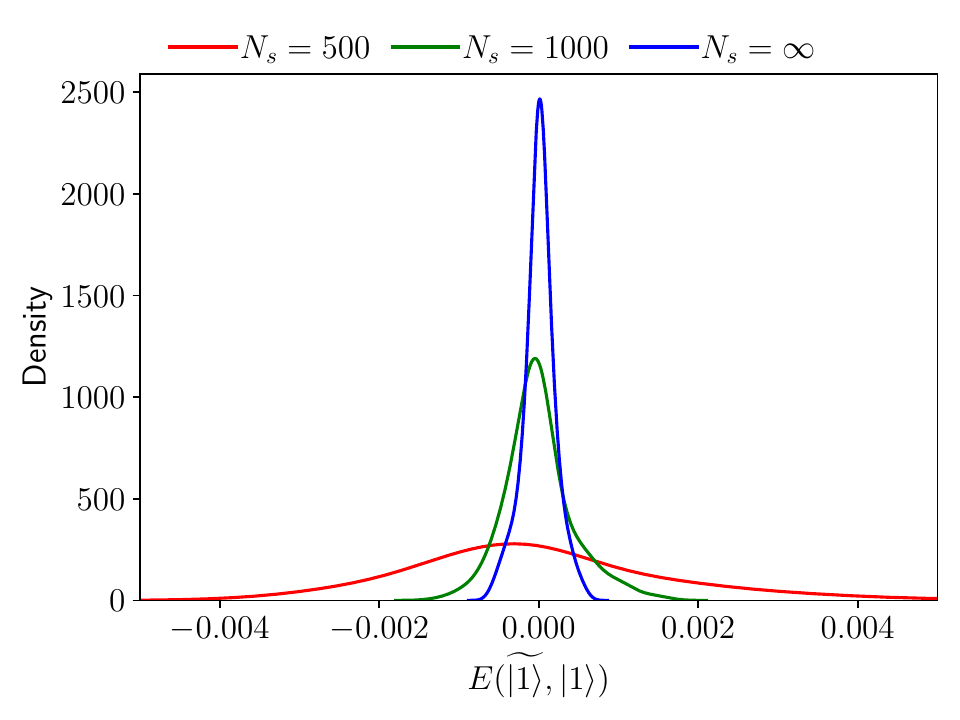} 
\end{subfigure}
\begin{subfigure}[h]{0.32\textwidth} 
\includegraphics[width=\textwidth]{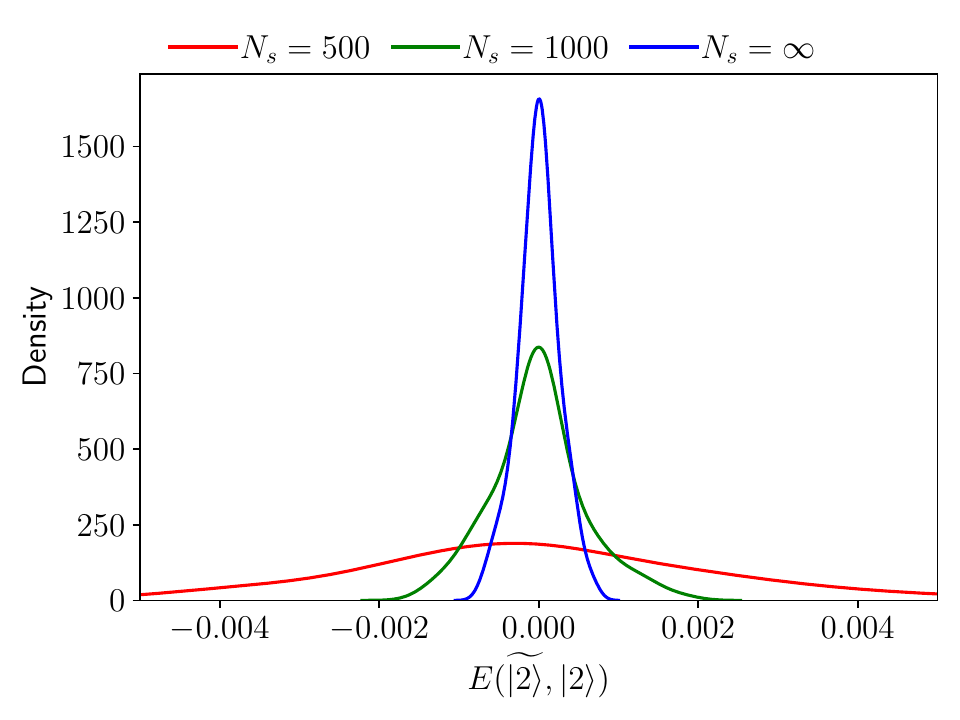} 
\end{subfigure}
\captionsetup{singlelinecheck=off,font=footnotesize}
\caption[]{(Top) Converged parameter distribution when performing \Added{stochastic} experiments and (Bottom) distribution of the error in the population when using different number of shots.  Note the parameter distributions are translated by the mean of the exact distributions.}
\label{fig:QC:Adaptive:PDF}
\end{figure}

Figure \ref{fig:QC:Adaptive:PDF} shows the distribution of the parameters and prediction errors whose mean and standard deviation are presents in Tab. \ref{table:Adaptive:Error}. Firstly, we see that in the absence of shot noise, the parameter mean is better estimated for all four parameters, whereas the presence of shot noise results in a shift in mean value and a larger spread. However, the distribution (and the statistical moment) of the error shows that \Added{iterative pulse extensions} lead to an order of magnitude reduction in error mean and variance (in the absence of shot noise) when compared to the fixed pulse parameters; the mean errors are also an order of magnitude smaller with comparable variances on the order of $10^{-3}$. This suggests that the BExD approach is able to reliably estimate the model parameters, even though the shot noise significantly affects the accuracy of the estimated parameters Furthermore, despite severe undersampling (using only 500 shots), the errors in the predictions from the calibrated model are well within typical bounds $\approx 99.5\%$ \cite{Wu2020} for several applications of NISQ testbeds.

\begin{table}[H]
\caption{Mean and standard deviation (in the brackets) of the error using \Added{iterative pulse extensions.}} \label{table:Adaptive:Error}
\centering
\begin{tabular}{c | c | c | c} 
\hline
Shots & $\ket{0}$ & $\ket{1}$ & $\ket{2}$\\
\hline
500 & 3.85\Added{$\times 10^{-5}$} (2.29\Added{$\times 10^{-3}$}) & -3.25\Added{$\times 10^{-5}$} (1.57\Added{$\times 10^{-3}$}) & -6.09\Added{$\times 10^{-6}$} (2.18\Added{$\times 10^{-3}$}) \\
1000 & 7.71\Added{$\times 10^{-6}$} (5.87\Added{$\times 10^{-4}$}) & 8.62\Added{$\times 10^{-7}$} (4.15\Added{$\times 10^{-4}$}) & -8.57\Added{$\times 10^{-6}$} (5.70\Added{$\times 10^{-4}$}) \\
$\infty$ & -1.76\Added{$\times 10^{-6}$} (2.76\Added{$\times 10^{-4}$}) & 4.87\Added{$\times 10^{-6}$} (1.93\Added{$\times 10^{-4}$}) & -3.11\Added{$\times 10^{-6}$} (2.57\Added{$\times 10^{-4}$}) \\
\hline
\end{tabular}
\end{table} 

\section{Conclusion}

The conventional techniques for quantum characterization, whether probabilistic or deterministic, rely on measurements from a set of experiments that are designed \emph{a-priori}. These measurements are then used to perform an offline calibration of the mathematical model to reconstruct the measurements from the experimental testbed. In most cases, the theoretical identifiability of the model parameters is neglected. Even when the identifiability conditions are known, manual design of experiments satisfying such conditions is not trivial. Hence, this work presented an automatic, online Bayesian framework for characterizing open \Added{qutrit} systems described by the Lindblad master equation. The approach automatically identifies optimal robust experiments that provide maximum information on the model parameters to be estimated. Furthermore, we proved the theoretical identifiability of the Schroedinger and Lindblad equations given a measured observable, and derived the sufficient conditions for the quantum state to establish the global, unique identifiability of the model parameters. 

We demonstrated our approach to identify the frequencies and decoherence parameters on a numerical test problem subjected to both parameter \Added{uncertainty} and shot noise. The characterization study demonstrated that different parameters exhibit different rates of convergence and are dependent on both the pulse parameterization and magnitude of shot noise. It was seen that accurate estimation of the frequencies required shortened pulse sequences whereas estimation of decoherence parameters required longer pulse sequences. We introduced an \Added{iterative pulse extension approach} to improve the convergence of all estimates. The results show that the Bayesian experiment design framework with \Added{iterative pulse extensions} is able to accurately estimate all four parameters in only 500 epochs (i.e. 500 different experiments). The approach is better able to estimate the distribution of the system frequencies than the decoherence parameters. Desipite this, the BExD approach yielded robust estimates of the parameters such that the mean prediction error and variances were on the order of $10^{-4}$. The results showed that the online BExD approach is able to design robust, optimal experiments to accurately characterize the quantum system described by the Lindblad master equation. Ongoing and future work includes experimentally demonstrating the technique on multi-qubit systems.

\section*{Acknowledgements}

The author is grateful to Dr. Yujin Cho at Lawrence Livermore National Laboratory for reviewing the manuscript and providing suggestions for improving it.
This work was funded in-part by the U.S. Department of Energy, Office of Advance Scientific Computing Research, Advanced Research in Quantum Computing program, project TEAM, award SCW-1683.1. This work was performed under the auspices of the U.S. Department of Energy by Lawrence Livermore National Laboratory under Contract DE-AC52-07NA27344. This is contribution LLNL-JRNL-858102.

\bibliographystyle{unsrt}
\bibliography{jobname}  

\appendix

\section{\Added{Robustness of Bayesian Experimental Design To Parameter Settings}} \label{sec:Appendix}

\begin{figure}[h]
\centering
%
\begin{subfigure}[h]{0.24\textwidth} 
\includegraphics[width=\textwidth]{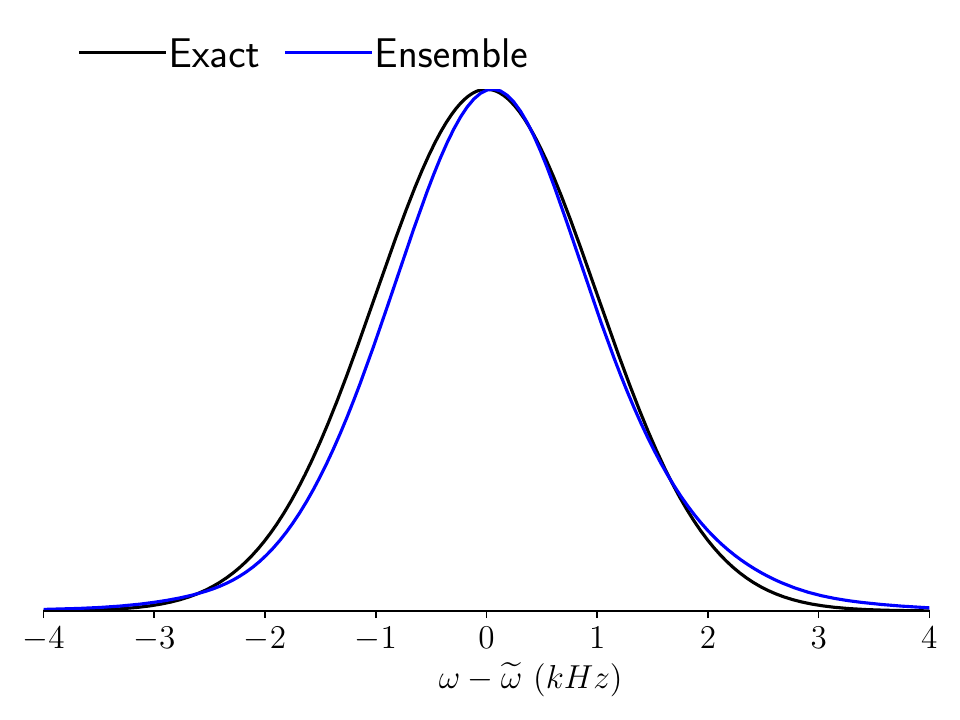} 
\end{subfigure}
\begin{subfigure}[h]{0.24\textwidth} 
    \includegraphics[width=\textwidth]{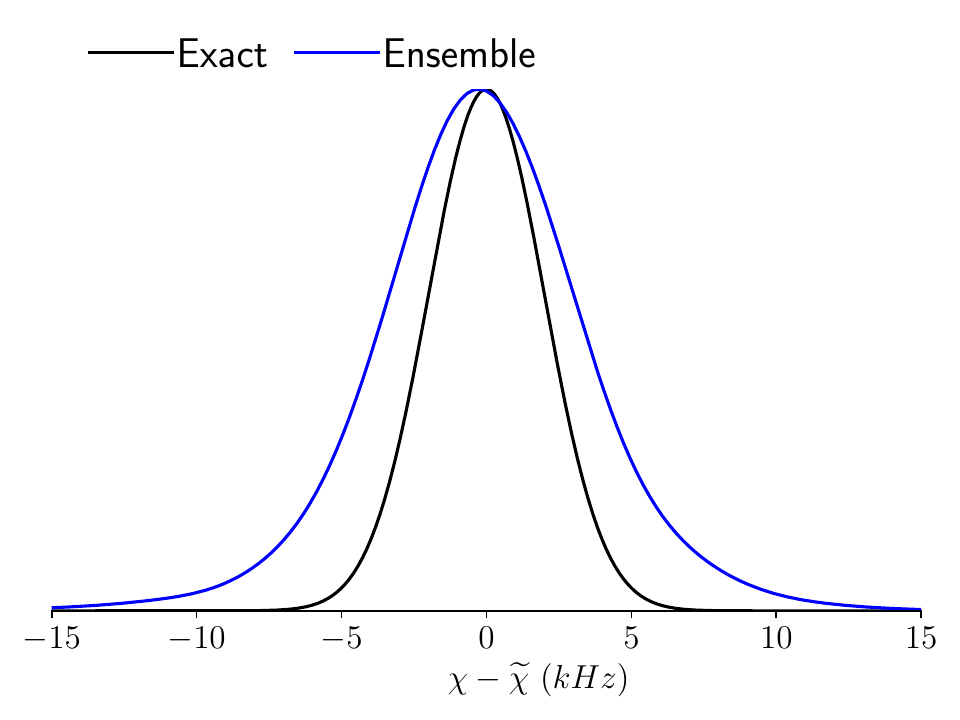} 
\end{subfigure}
\begin{subfigure}[h]{0.24\textwidth} 
    \includegraphics[width=\textwidth]{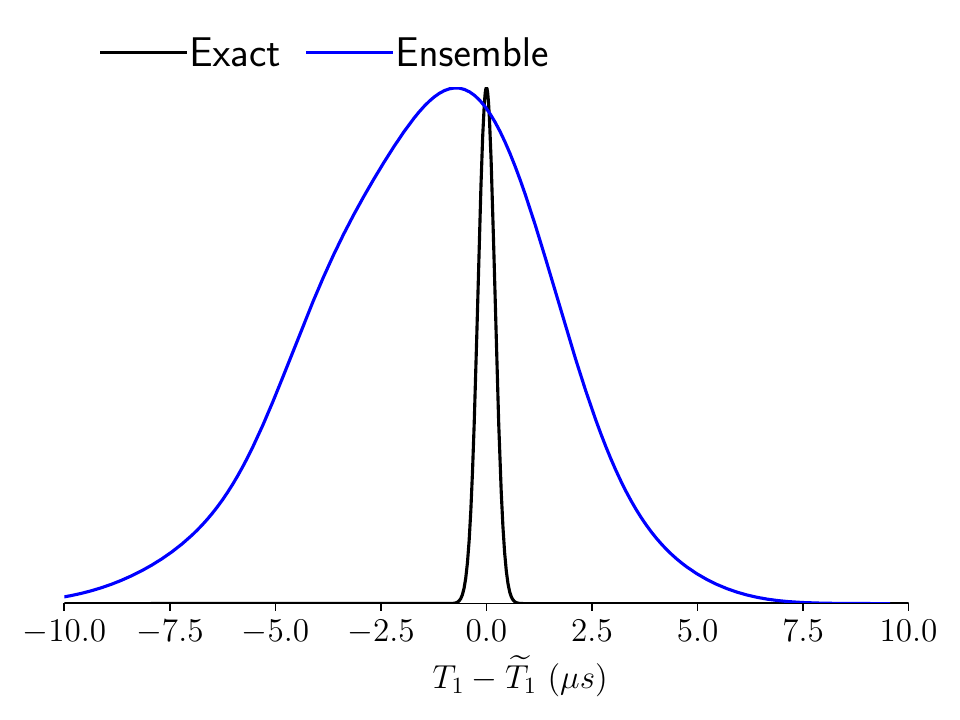} 
\end{subfigure}
\begin{subfigure}[h]{0.24\textwidth} 
    \includegraphics[width=\textwidth]{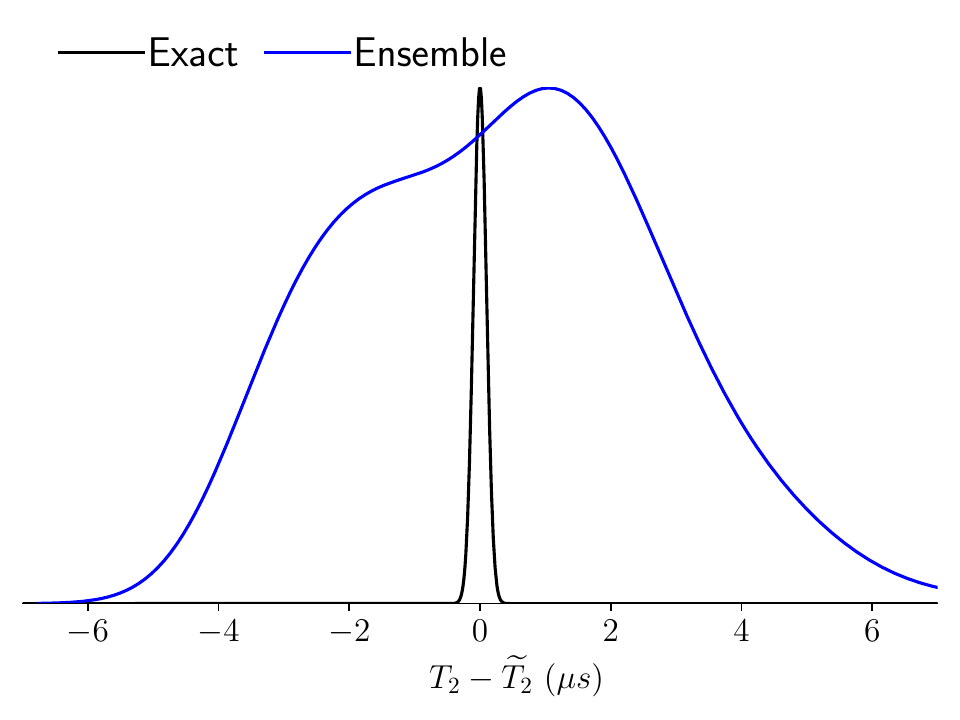} 
\end{subfigure}
\begin{subfigure}[h]{0.32\textwidth} 
\includegraphics[width=\textwidth]{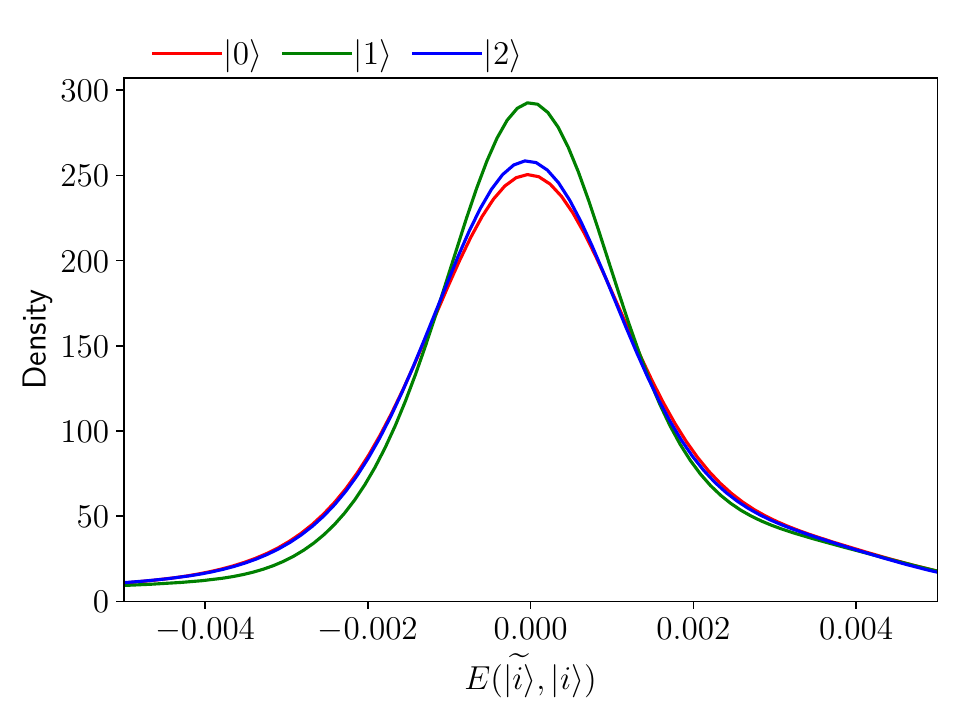} 
\end{subfigure}
%
%
\captionsetup{singlelinecheck=off,font=footnotesize}
\caption[]{(Top) Converged parameter distribution over an emsemble of six runs with when performing \Added{stochastic} experiments with parameter uncertainty and (Bottom) distribution of the error in the population.  Note the parameter distributions are translated by the mean of the exact distributions.}
\label{fig:QC:Adaptive:PDF:Ensemble}
\end{figure}

Figure \ref{fig:QC:Adaptive:PDF:Ensemble} shows the distribution of parameters and prediction errors from an ensemble (with different random initialization of 2000 particles) of six runs using the same control parameterization as those in Section \ref{subsec:AdaptivePulse}. Here, the experiments are considered to be stochastic with parameter uncertainty but without shot noise. The distributions are similar to those in Fig. \ref{fig:QC:Adaptive:PDF} and shows the estimation is robust to different initialization strategies. 

\begin{figure}[h]
\centering
%
\begin{subfigure}[h]{0.24\textwidth} 
\includegraphics[width=\textwidth]{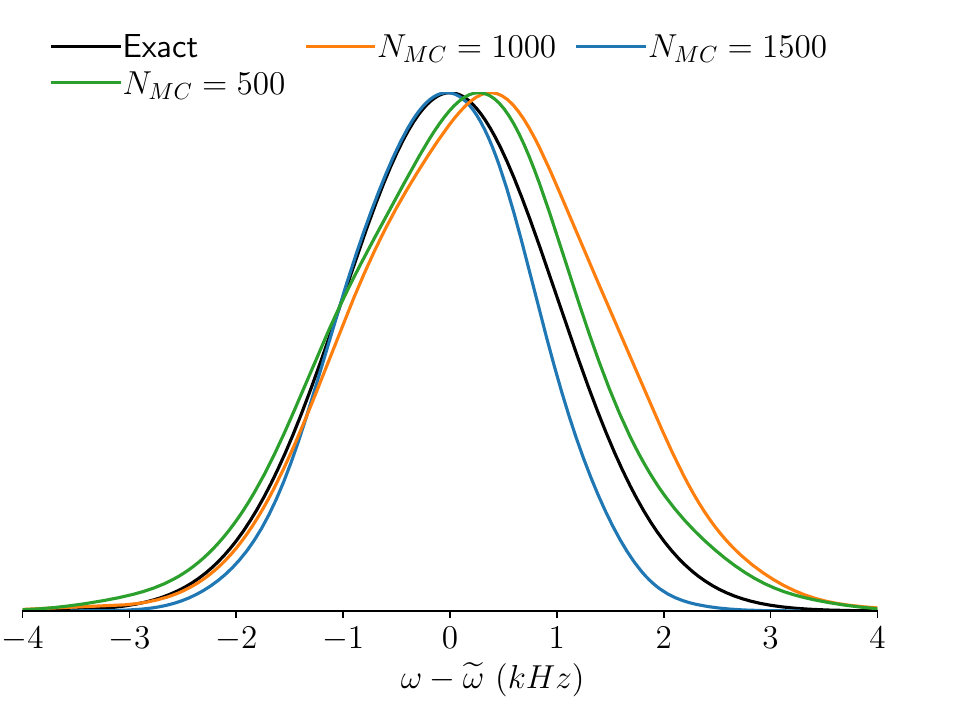} 
\end{subfigure}
\begin{subfigure}[h]{0.24\textwidth} 
    \includegraphics[width=\textwidth]{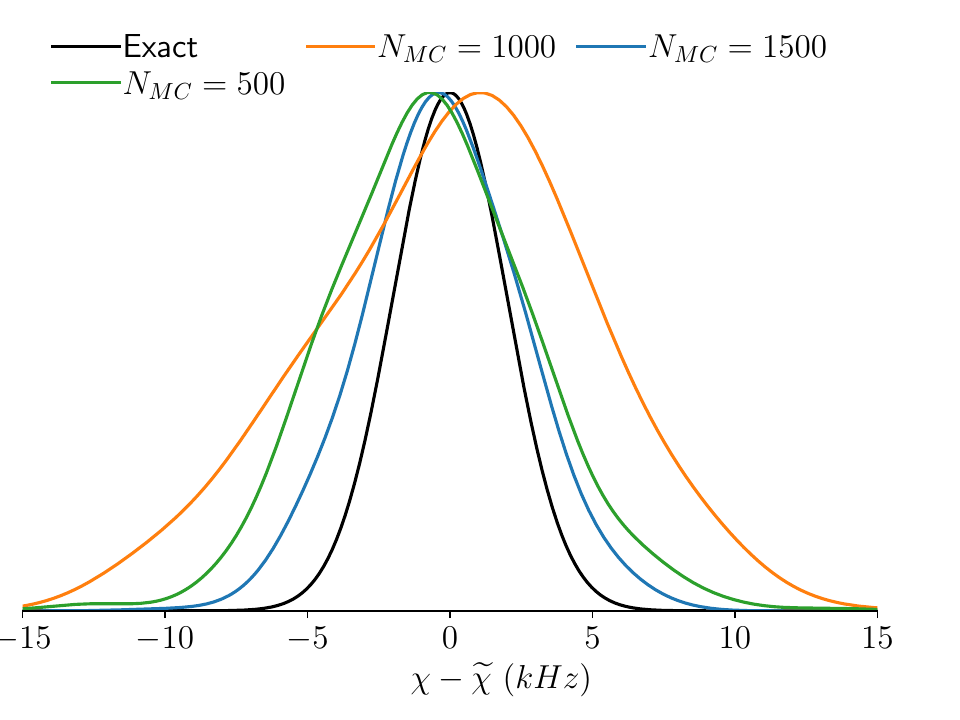} 
\end{subfigure}
\begin{subfigure}[h]{0.24\textwidth} 
    \includegraphics[width=\textwidth]{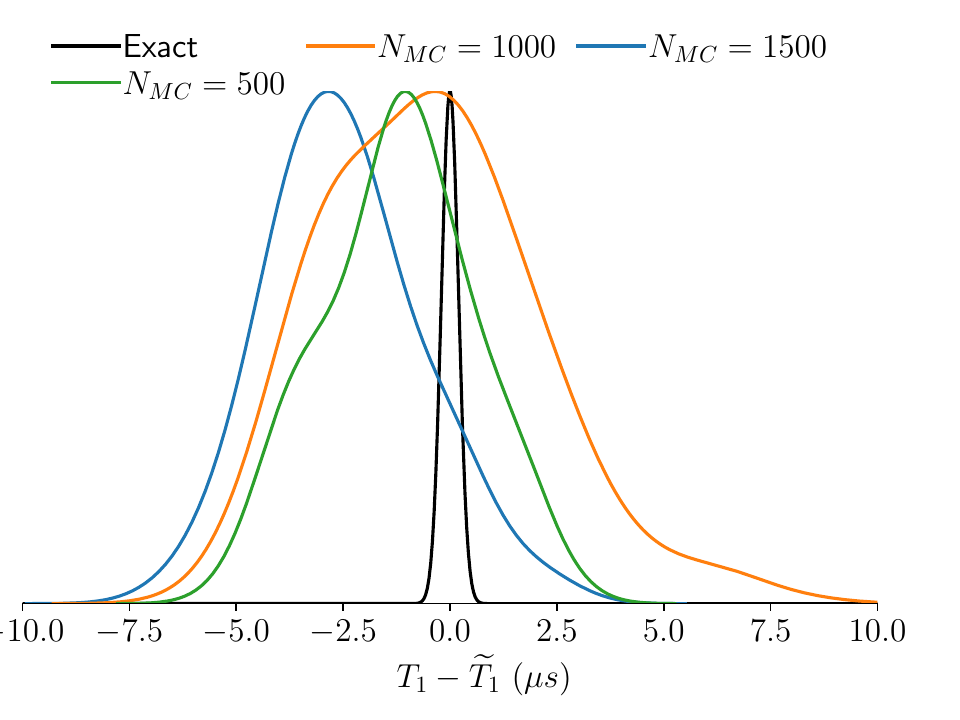} 
\end{subfigure}
\begin{subfigure}[h]{0.24\textwidth} 
    \includegraphics[width=\textwidth]{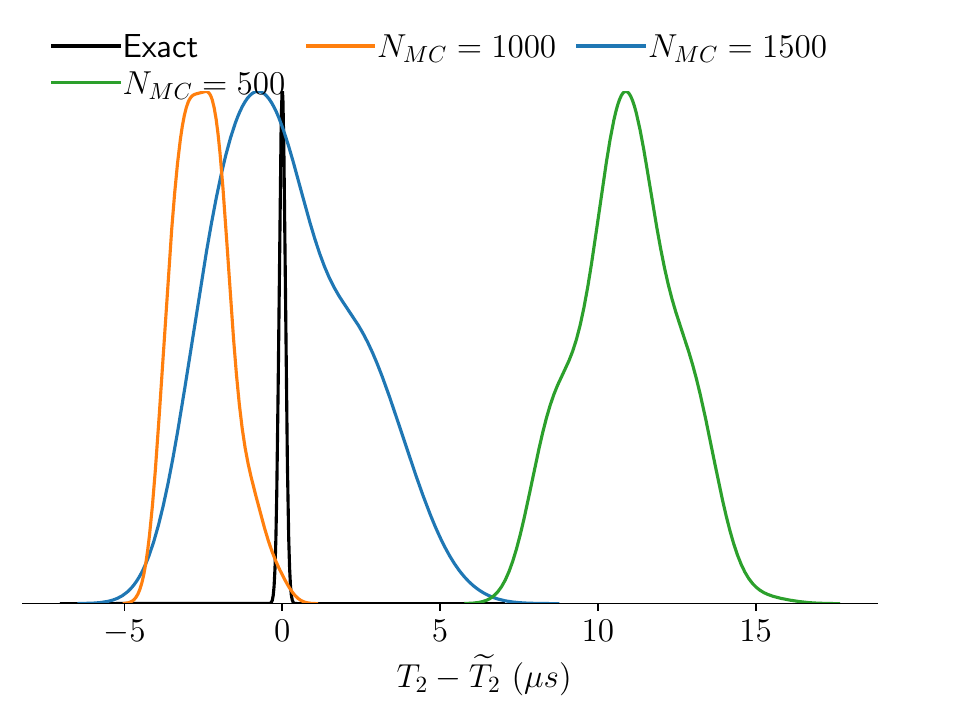} 
\end{subfigure}
\begin{subfigure}[h]{0.32\textwidth} 
\includegraphics[width=\textwidth]{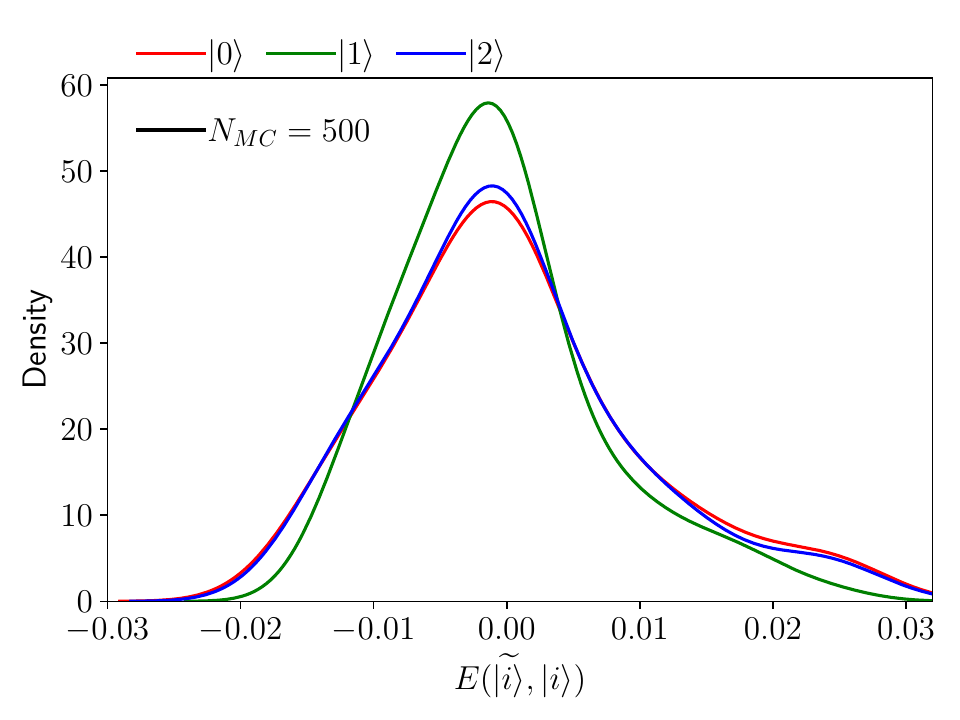} 
\end{subfigure}
\begin{subfigure}[h]{0.32\textwidth} 
\includegraphics[width=\textwidth]{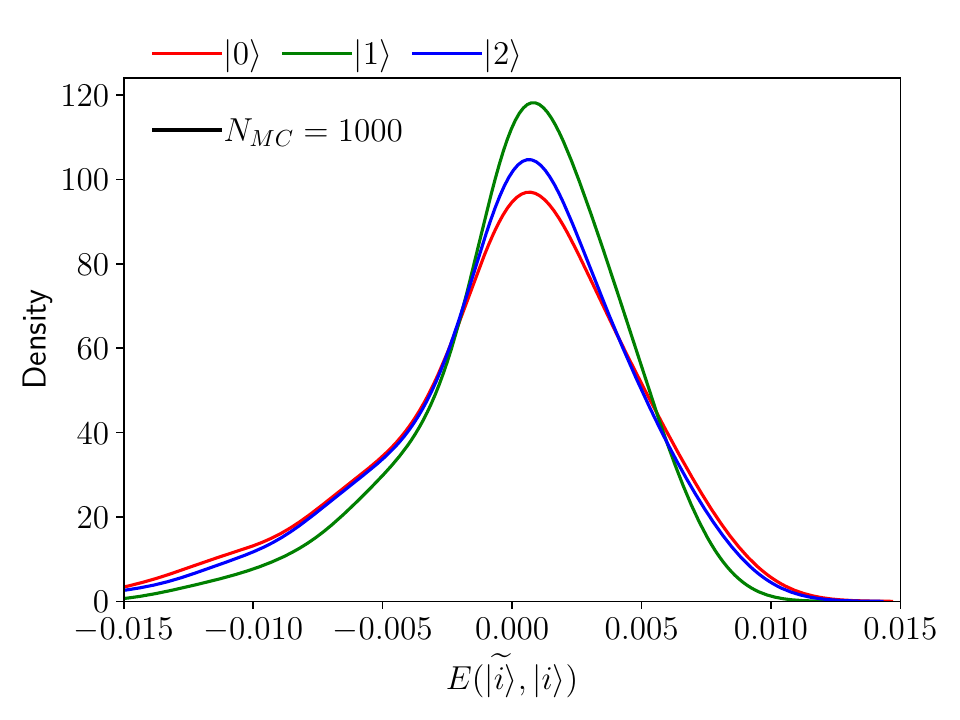} 
\end{subfigure}
\begin{subfigure}[h]{0.32\textwidth} 
\includegraphics[width=\textwidth]{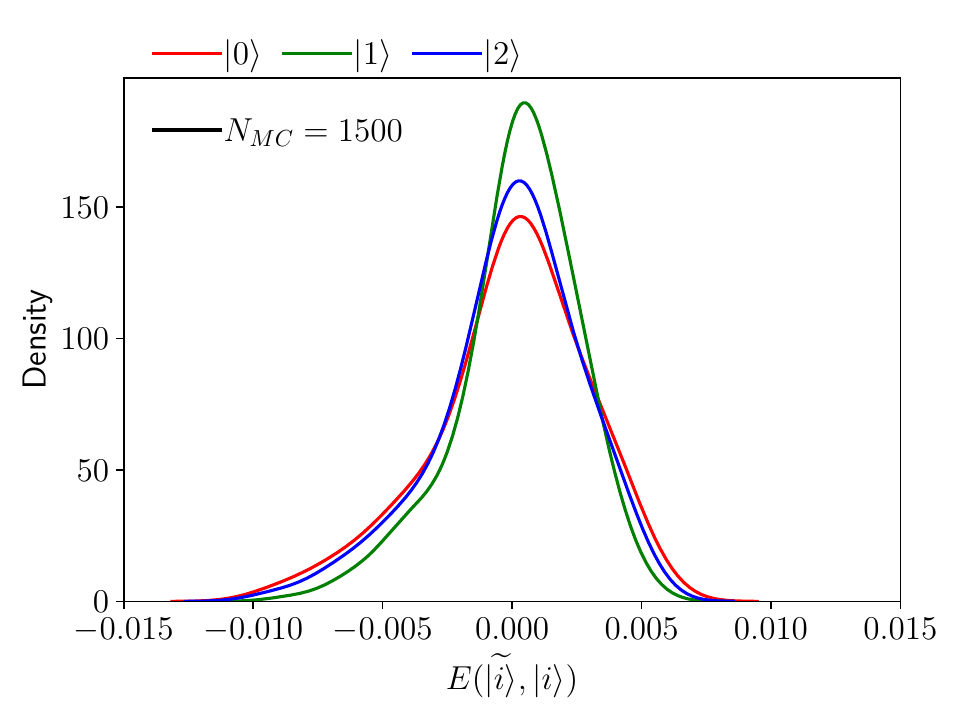} 
\end{subfigure}
\captionsetup{singlelinecheck=off,font=footnotesize}
\caption[]{(Top) Converged parameter distribution when performing \Added{stochastic} experiments with parameter uncertainty and (Bottom) distribution of the error in the population when using different number of Monte Carlo particles.  Note the parameter distributions are translated by the mean of the exact distributions.}
\label{fig:QC:Adaptive:PDF:MCSamples}
\end{figure}

Figure \ref{fig:QC:Adaptive:PDF:MCSamples} shows the distribution of parameters and prediction errors when using different number of Monte Carlo particles. Here, the experiments are again considered to be stochastic with parameter uncertainty but without shot noise. The distributions show that the frequency parameters are estimated more accurately and the decoherence times, with only a few number of samples. The under-resolved parameterization of the distribution (\eqref{BExD:PDF}) and estimation of expectations by Monte Carlo sampling (\eqref{BExD:MC}), leads to this decrease in estimation accuracy, which in turn leads to an increase in prediction errors. However, this shows that the BExD approach does not require an overwhelming number of particle but a sufficient number to cover the parameter space. 

\end{document}